\documentclass[fleqn,usenatbib]{mnras}

\usepackage{newtxtext,newtxmath}
\usepackage[T1]{fontenc}

\DeclareRobustCommand{\VAN}[3]{#2}
\let\VANthebibliography\thebibliography
\def\thebibliography{\DeclareRobustCommand{\VAN}[3]{##3}\VANthebibliography}

\usepackage{graphicx}	% Including figure files
\usepackage{amsmath}	% Advanced maths commands
\usepackage{comment}
\usepackage{multicol}
\usepackage{multirow}
\usepackage{lscape}
\usepackage{threeparttable}
\usepackage{xcolor}

\def\arcsec{$^{\prime\prime}$}
\def\arcmin{$^{\prime}$}

\newcommand{\NtwoH}{N$_{2}$H$^{+}$}

\newcommand{\kms}{km~s$^{-1}$}

\newcommand{\revt}[1]{{\color{black}#1}} 
\newcommand{\revtt}[1]{{\color{black}#1}} 
\title[Dynamics in Star-forming Cores (DiSCo)]{Dynamics in Star-forming Cores (DiSCo): Project Overview and the First Look toward the B1 and NGC\,1333 Regions in Perseus}

\author[C.-Y. Chen et al.]{
Che-Yu Chen,$^{1,2}$\thanks{E-mail: cheyu.c@gmail.com}
Rachel Friesen,$^{3}$
Jialu Li,$^{4}$
Anika Schmiedeke,$^{5,6}$
David Frayer,$^{6}$
Zhi-Yun Li,$^{2}$
\newauthor John Tobin,$^{7}$
Leslie W. Looney,$^{8}$
Stella Offner,$^{9}$
Lee G. Mundy,$^{4}$
Andrew I. Harris,$^{4}$
Sarah Church,$^{10}$
\newauthor Eve C. Ostriker,$^{11}$
Jaime E. Pineda,$^{5}$
Tien-Hao Hsieh,$^{5}$
and Ka Ho Lam$^{2}$
\\\\
$^{1}$Lawrence Livermore National Laboratory, Livermore, CA 94550, USA\\
$^{2}$Department of Astronomy, University of Virginia, Charlottesville, VA 22904, USA\\
$^{3}$Department of Astronomy \& Astrophysics, University of Toronto, Toronto, ON M5S 3H4, Canada\\
$^{4}$Department of Astronomy, University of Maryland, College Park 20742, MD, USA\\
$^{5}$Max-Planck-Institut f\"ur extraterrestrische Physik, Giessenbachstrasse 1, D-85748 Garching, Germany\\
$^{6}$Green Bank Observatory, Green Bank, WV 24944, USA\\
$^{7}$National Radio Astronomy Observatory, Charlottesville, VA 22904, USA\\
$^{8}$Department of Astronomy, University of Illinois, Urbana, IL 61801, USA\\
$^{9}$Department of Astronomy, University of Texas, Austin, TX 78712, USA\\
$^{10}$Physics Department, Stanford University, Stanford, CA 94305, USA\\
$^{11}$Department of Astrophysical Sciences, Princeton University, Princeton, NJ 08544, USA\\
}
\date{Accepted XXX. Received YYY; in original form ZZZ}

\pubyear{2023}

\begin{document}
\label{firstpage}
\pagerange{\pageref{firstpage}--\pageref{lastpage}}
\maketitle

\begin{abstract}
% now 190 words
The internal velocity structure within dense gaseous cores plays a crucial role in providing the initial conditions for star formation in molecular clouds. However, the kinematic properties of dense gas at core scales ($\sim 0.01-0.1$\,pc) has not been extensively characterized because of instrument limitations until the unique capabilities of GBT-Argus became available.
The ongoing GBT-Argus Large Program, Dynamics in Star-forming Cores (DiSCo)
thus aims to investigate the origin and distribution of angular momenta of star-forming cores. 
DiSCo will survey all starless cores and Class 0 protostellar cores in the Perseus molecular complex down to $\sim 0.01$\,pc scales with $<0.05$\,\kms\ velocity resolution using the dense gas tracer \NtwoH.
%With the full dataset, DiSCo will address a key open question in star formation: the origin and distribution of angular momenta of star-forming cores. 
Here, we present the first datasets from DiSCo toward the B1 and NGC\,1333 regions in Perseus. Our results 
%generally agree with recent theoretical and observational work 
suggest
that a dense core's internal velocity structure has little correlation with other core-scale properties, indicating these gas motions
may be originated externally from cloud-scale turbulence. 
These first datasets also reaffirm the ability of GBT-Argus for studying dense core velocity structure and provided an empirical basis for future studies that address the angular momentum problem with a statistically broad sample.
\end{abstract}


%%%%%%%%%%%%%%%%%%%%%%%%%%%%%%%%%%%%%%%%%%%%%%%%%%

%%%%%%%%%%%%%%%%% BODY OF PAPER %%%%%%%%%%%%%%%%%%

\section{Introduction}
\label{sec:intro}

%\begin{comment}
In molecular clouds (MCs), multi-scale supersonic flows compress material to initiate the creation of overdense regions, which may shrink to become prestellar cores and then collapse gravitationally to form protostellar systems \citep{Shu1987}. 
Turbulence is therefore considered one of the key agents affecting the dynamics of the star forming process in MCs, in combination with magnetic fields and gravity, at all physical scales and throughout different evolutionary stages \citep{MO2007,Girichidis_SFreview_2020}. 
There have been several observational projects aiming to resolve the gas dynamic properties at larger (MCs; e.g.,~the CARMA Large Area Star Formation Survey (CLASSy),~\citealt{Storm_CLASSy_2014}; the Green Bank Ammonia Survey (GAS),~\citealt{GAS_2017}) and smaller (protostellar disks; e.g.,~the VLA/ALMA Nascent Disk and Multiplicity (VANDAM) Survey,~\citealt{Tobin_outflow_2016}; the Mass Assembly of Stellar Systems and their Evolution with the SMA (MASSES) survey,~\citealt{Stephens_MASSES_2018}) scales.
These observations revealed features that generally agree with the picture that multi-scale turbulence is dynamically significant during the star formation process, including misalignment between outflow direction and core-scale magnetic field orientation \citep{Hull+14, Xu_Offner_2022}, misalignment between disk orientation and filament structure \citep{Stephens+17}, the correlation between dense core location and turbulence structure \citep{HopeGAS19}, and the misalignment between core elongation and local magnetic field orientation \citep{Chen_PlanckGAS_2020,Pandhi_Friesen_coreAlign_2023}.
On the other hand, the velocity information within dense cores,
especially on the crucial scales of $0.01-0.05$\,pc that contain the bulk of the material destined to be incorporated into disks and subsequently accreted onto the star or planets, 
remains less well-investigated. 
%because this scale ($\sim 7-60$\arcsec{} in nearby clouds with distance $d\sim 150-300$\,pc) is generally beyond the resolution of single-dish telescopes for low-J molecular transitions that trace these cold dense gas, while most of the interferometers cannot cover such a large spatial area with the necessary velocity resolution within a reasonable amount of time.

As the immediate precursors of protostars, dense molecular cores possess critical information about the physical properties at the earliest stage of star formation.
%The observed velocity gradient is commonly used, when present, to estimate (or used for upper limits of) 
In particular, rotation within dense cores is important in the evolution leading to the creation of protostellar systems \citep[see review in][]{Li2014PPVI}. 
The angular momentum of star-forming cores is hence
%a critical parameter in protostellar evolution, but its origin is not well understood (see the review in Li et al. 2014). 
a critical quantity in shaping the outcome of core collapse: whether a single star or multiple system is formed \citep{OffnerPPVII}, and whether a large or small disk is produced \citep{Li2014PPVI}. 
It is also important for determining whether the angular momentum of core material is conserved during the collapse into a star-disk system or whether it is significantly reduced; the latter would point to an efficient mechanism of angular momentum removal, such as magnetic braking \citep[see review in][]{Li2014PPVI}.

%Observationally determining the core-scale angular momentum is a pre-requisite for understanding the formation and diversity of protostellar disks and binary/multiple systems. 
While there have been several proposals on the origin of core-scale angular momentum, including the cloud-scale turbulence \citep{BB_turb_2000,CO18}, the dynamic interaction between local clumps \citep{Kuznetsova_AngMom_2019}, and filament fragmentation \citep{Misugi_AngMom_2019}, 
it is still not well understood through observations, 
because this scale ($\sim 7-60$\arcsec{} in nearby clouds with distance $d\sim 150-300$\,pc) is generally beyond the resolution of single-dish telescopes for low-J molecular transitions in wavelengths~$\sim 3$\,mm regime that trace the cold dense gas, while most of the interferometers cannot cover such a large spatial area with the necessary velocity resolution within a reasonable amount of time.
Because of these constraints,
linear fitting is generally applied to observed velocity gradients across cores, regardless of the complex nature of the velocity field. It is assumed that rigid-body rotation applies and that the angular speed is roughly the gradient of line-of-sight velocity \citep{Goodman1993}.

It is known that some dense cores show clear gradients in line-of-sight velocities, while others have a relatively random velocity field \citep[see e.g.,][]{Goodman1993,Caselli2002}. 
Previous observations \citep[e.g.,][]{Goodman1993,Caselli2002,Pirogov_2003,XChen_2007,Tobin_2011,HChen2019b,Pandhi_Friesen_coreAlign_2023} as well as fully-3D MHD simulations \citep{CO15,CO18} have found a power-law relationship, $J \sim R^{1.5}$, between the \revt{total} specific angular momentum $J$, defined as the \revt{total} angular momentum $L$ divided by mass $M$, and radius $R$ for dense cores/clumps with radii $\sim ~0.005-10$~pc \citep[see e.g.,][]{Li2014PPVI,CO18}.
%though observations remain relatively sparse
\revt{
More recently, higher angular resolution maps enabled the derivation of the specific angular momentum radial profile, $j(r)$, which shows that this radial dependence is closer to $j(r) \propto r^{1.8}$ \citep[e.g.,][]{Pineda_jr_2019,Gaudel_jr_2020}.
}
The $J-R$ correlation over such a large range of spatial scales suggests that either gas motion in cores originates at scales much larger than the core size, or the observed rotation-like features arise from sampling of turbulence at a range of scales \citep{BB_turb_2000}.

However, observational statistics are quite poor on the crucial sub-core scales of $\sim 0.01-0.05$~pc due to the aforementioned instrumental constraints.
%\todo{[mention GAS somewhere: a precursor to this, which investigated many similar themes about cores]}
For example, the recent large-scale NH$_3$ survey GAS \citep{GAS_2017} was able to reveal dense gas dynamics in all the northern Gould Belt clouds with $A_V > 7$\,mag, but has very limited resolution at sub-pc scale due to the $30$\arcsec{} beam. 
The previous interferometer survey CLASSy \citep{Storm_CLASSy_2014} was able to resolve \NtwoH{} down to $\sim 7$\arcsec{} scale in three nearby star-forming clouds, but it lacks the necessary spectral resolution for the fine velocity structure within cores. More discussions on recent observational surveys can be found in the review by \cite{Pineda_PPVII}.

Here, we present the first results from the Dynamics in Star-forming Cores (DiSCo) Survey with the recently commissioned {\it Argus} focal plane array on the Green Bank Telescope (GBT).
Unlike interferometers which have limited spatial frequency coverage, GBT-Argus has full spatial sampling for both compact and extended emission.
With the unique combination of angular and velocity resolution, full spatial-scale recovery, and imaging speed, GBT-{\it Argus} is ideal to survey a large sample of dense cores in nearby star-forming MCs with high resolution and sensitivity.
\revt{Built on a successful pilot study \citep{GBTcore19},}
the DiSCo survey aims to provide an unprecedented observational view of how the original core material forms and falls into the protostellar system.
%The best way to provide new insight to the above unanswered questions is by surveying a large sample of dense cores in nearby star-forming MCs with high resolution and sensitivity. 

The outline of this paper is as follows. We introduce the DiSCo project in Section~\ref{sec:project}, including detailed descriptions on target selection (Section~\ref{sec:target}), observations (Section~\ref{sec:obs}), and our data reduction and spectral line fitting routines (Section~\ref{sec:data}). 
The cleaned results are presented in Section~\ref{sec:result}, where we discuss our core identification method (Section~\ref{sec:core}) and the derived core-scale angular momentum (Section~\ref{sec:JR}). Further discussions are presented in Section~\ref{sec:rotation}, and we summarize our results in Section~\ref{sec:sum}.
%\end{comment}

\section{Project overview}
\label{sec:project}

\begin{figure*}
	\includegraphics[width=0.8\textwidth]{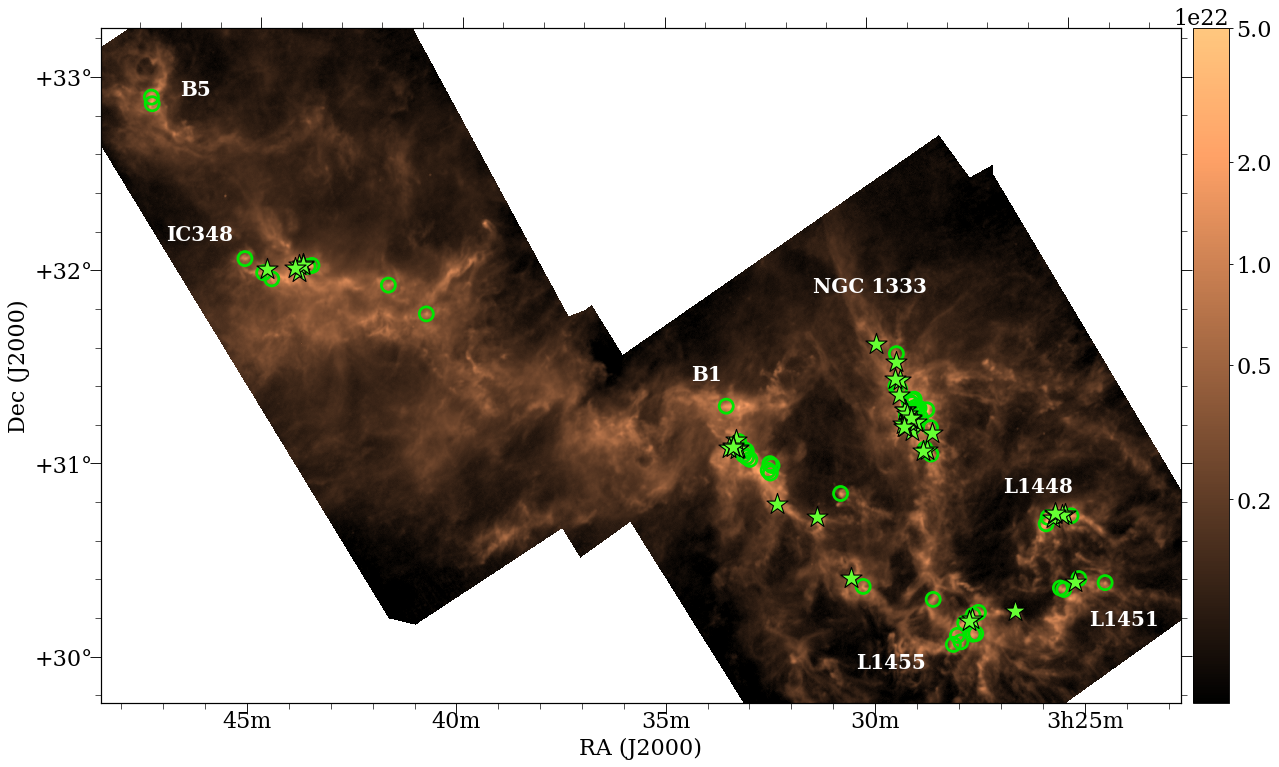}
    \vspace{-0.05in}
    \caption{{\it Herschel} H$_2$ column density map (in cm$^{-2}$) of the actively star-forming Perseus molecular cloud over-plotted with the $108$ Class~0 protostellar ({\it stars}) and starless ({\it circles}) targets of the DiSCo survey. }
    \label{fig:Herschel}
\end{figure*}

DiSCo is a GBT large program (GBT20A-322, PI:~C.-Y.\,Chen) aiming to systematically characterize the internal velocity structures of young dense cores in star-forming regions.
The Perseus Molecular Cloud was chosen because of its relatively-close distance \citep[$d \approx 300$\,pc;][]{Zucker+2019}, a wide range of levels of star formation activity \citep[e.g.,][]{Storm_CLASSy_2014}, and excellent ancillary data \citep[see e.g.,][]{Tobin_VANDAM_2016,Stephens_MASSES_2018,HopeGAS19,Hsieh_ALMA_2019}.

\subsection{Target Selection}
\label{sec:target}

We construct a target catalog containing a sample of 108 cores in Perseus.
Based on previous \NtwoH{} observations \citep[e.g.,][]{Storm_CLASSy_2014} and our pilot studies \citep{GBTcore19}, we decided to focus on Class\,0 protostellar cores because cores at later evolutionary stages (Class\,I and beyond) tend to have less \NtwoH{} emission \citep[see e.g.,][]{Hsieh_ALMA_2019} . 
We used the catalog from the VANDAM Survey \citep{Tobin_VANDAM_2016} for the Class 0 protostars in Perseus. 
Among the 45 Class\,0 protostars in Perseus, 21 of them are single stars, while the remaining 24 are either a \revt{compact} multiple system itself, or part of a multiple system.
\revt{These are all included in the DiSCo project.}
There are 8 protostars in B1 and 21 in NGC\,1333.

For starless cores, we cross-compared and combined 3 separate catalogs to achieve a complete census. Most of the cores are based on the \NtwoH{} dendrogram leaves identified in CLASSy \citep{Storm_CLASSy_2014}) which provided the most direct comparison to the DiSCo project. However, CLASSy only covers the densest areas in three of the six sub-regions (B1, NGC 1333, and L1451) in Perseus (see Fig.~\ref{fig:Herschel}). Since 1-to-1 correlation is often found in \NtwoH{} and NH$_3$ cores \citep[e.g.,][]{Johnstone_NH3_2010}, we also consider NH$_3$ dendrogram leaves identified in GAS \citep{GAS_2017}) which covers all $A_V > 7$\,mag sub-regions in Perseus. To make sure that we include isolated cores as well, dense cores identified in JCMT-SCUBA 850\,$\mu$m observation with bright \NtwoH{} emission detected by the 30\,m IRAM telescope are also included \citep{Kirk_Perseus_2007,Johnstone_NH3_2010}. 

From the dendrogram analysis, we only consider cores with minor axes $\gtrsim 20$\arcsec{} (i.e. at least two beams across the narrowest region within the core) and aspect ratio $\lesssim 2.5$ (to avoid highly elongated structures like filaments). To limit the observational time to a reasonable amount, we select only bright cores with peak \NtwoH{} intensities $\gtrsim 2.0$\,K. We also checked the YSO catalogs from \cite{Jorgensen_Perseus_2007} and the VANDAM survey to make sure these cores are starless.
63 starless cores are identified as DiSCo targets, among which there are 13 starless cores in B1 and 24 in NGC\,1333.
Figure~\ref{fig:Herschel} shows an annotated {\it Herschel} column density map \citep[\revt{resolution $36.3$\arcsec};][]{Andre_HGBS_2010,Pezzuto_Herschel_2021} illustrating the locations of our targets.
%These selected dense cores have effective radii $\sim 20$\arcsec (about $0.03$~pc at $d\approx 300$~pc), which is around the critcal scale in studying the evolution of angular momentum within star-forming cores where gravitational collapse normally happens.
%Also, the targets were chosen to cover different evolutionary stages (a mix of starless and protostellar cores) and ambient environments (cluster neighborhood versus relatively isolated cores). 

\subsection{Observations}
\label{sec:obs}

{\it Argus} is a 16-pixel focal plane array operating in the $74-116$~GHz range on the GBT \citep{Argus2014}, which is designed for efficient and sensitive large-area mapping. 
%For a spectral line observation over a region much larger than the telescope beam size, the mapping speed scales as $\sim n_\mathrm{pix}/(\Delta T_\mathrm{min})^2$, where $n_\mathrm{pix}$ is the number of receiving pixels and $\Delta T_\mathrm{min}$ is the line detection sensitivity. 
As GBT provides a beam size of $\sim 9$\arcsec{} at 90~GHz, the $4\times 4$ {\it Argus} array with each receiving pixel separated by 30.4\arcsec{} on the sky is able to significantly improve the mapping speed. 
Also, {\it Argus} has a low system temperature by using advanced Monolithic Millimeter-wave Integrated Circuit (MMIC) technology,
%\citep{MMIC_2009}, 
which gives receiver noise temperatures of less than 53\,K per pixel.

The observations presented here were conducted using the GBT from semester 2020A through 2022A with the on-the-fly (OTF) method (see \citealt{Mangum_OTF_2007} and references therein). 
The total observing time is about 116\,hr.
We started each session with observations of a calibrator to adjust the telescope surface for thermal corrections, to determine pointing corrections and receiver focus. 
We used the {\it Argus} receiver to map \NtwoH{} J=1-0 emission with the VEGAS backend, which was configured to a rest frequency of~93173.704~MHz for \NtwoH{} using mode~6 with 187.5 MHz of bandwidth and 1.43 kHz ($\sim 0.0046$\,\kms) spectral resolution. 
System temperature calibration for all 16 receiver pixels was done before observing the science targets, and we performed calibration, pointing, and focus scans every 30-50 minutes depending on the weather.
%We mapped the science targets in RA and DEC scan directions. 
%The on-source integration time was determined using previous \NtwoH{} data in Perseus from the CARMA Large Area Star Formation Survey \citep[CLASSy;][]{2014ApJ...794..165S} to reach our requested sensitivity $\sim 0.05$~K.
Based on our previous pilot study, we set the integration time per beam for the 16 {\it Argus} beams to be $25-27$ seconds in order to achieve our desired sensitivity of $\sim 0.05$\,K \revt{in brightness temperature}.
%The data were taken every 2~seconds with map scan rates of about 0.92\arcsec{} per second; this led to angular sampling of about 1.8\arcsec{} per sample, which was about $4-5$ times less than the expected angular size of the beam. 
We keep the sampling rate $\sim 1.2-1.8$\arcsec{} per sample so that it is less than $1/3$ of the beam size ($\sim 9$\arcsec). This means the integration time (data output rate) is either 1 second with a map scan rate of $\sim 1.2-1.8$\arcsec{} per second ($t_{\rm int} = 1$) or $t_{\rm int} = 2$~second with a scan rate $\sim 0.6-0.9$\arcsec{} per second.
Frequency switching was used with offsets of $-12.5$ and $+12.5$~MHz. 
The calibrated beam size is 9.4\arcsec{} for \NtwoH{} J = 1-0 emission. 
%\todo{LWL: Give amp calibration uncertainties, then say statistically only is considered afterwards.}

\subsection{Data reduction and spectral fitting}
\label{sec:data}

{\it Argus} uses the chopper wheel method for calibration, which is the standard procedure in mm and sub-mm spectral line observation \citep{Kutner_calibration_1981}. The data is thus calibrated in ${T_a}^*$ scale, which corrects for atmospheric attenuation, resistive losses and rearward spillover and scattering. The GBT weather database returns the atmospheric temperature and opacity information. 
The main-beam efficiency of {\it Argus} is adopted to be $\eta_\mathrm{MB} = 0.505$ for \NtwoH{} at 93.17~GHz with a typical measurement uncertainty of $\sim 20\%$,
%the amplitude calibration uncertainty is estimated to be ~10\%, 
and the flux uncertainties adopted in this work are statistical only.
We refer the readers to the GBT-{\it Argus} calibration memo, \cite{Frayer_Argus_2019}, for more details.

We performed standard calibration using \texttt{GBTIDL} including baseline subtraction, and \texttt{gbtgridder} was used to make data cubes with pixel size 2\arcsec~$\times$~2\arcsec{} from maps per frequency channel using a Gaussian kernel. 
When re-gridding the data, we chose to combine 5 channels for \NtwoH{} to reach velocity resolution $\approx 0.023$~\kms. The sensitivity of our final \NtwoH{} maps is $\sim 0.2-0.25$~K.
We used the Python package \texttt{PySpecKit} \citep{PySpecKit2011,pyspeckit2022} for spectral fitting, which simultaneously fits the 7 hyperfine lines of \NtwoH{} J=1-0 and returns the fitted centroid velocity, linewidth, excitation temperature, and optical depth. We adopted a signal-to-noise cut to peak line intensity of $S/N > 5$ for the \NtwoH{} data when performing the fitting, and a $S/N > 2$ mask when generating \NtwoH{} moment~0 maps, 
\revt{which are constructed by integrating emission from all 7 hyperfine structures of \NtwoH}
(see Figs.~\ref{fig:B1result}$-$\ref{fig:N1333result}).
\vspace{-.2in}

\section{Results}
\label{sec:result}
\vspace{-.05in}

%\todo{LWL: Something about uncertainties needs to be mentioned.}
\begin{figure*}
	\includegraphics[width=\textwidth]{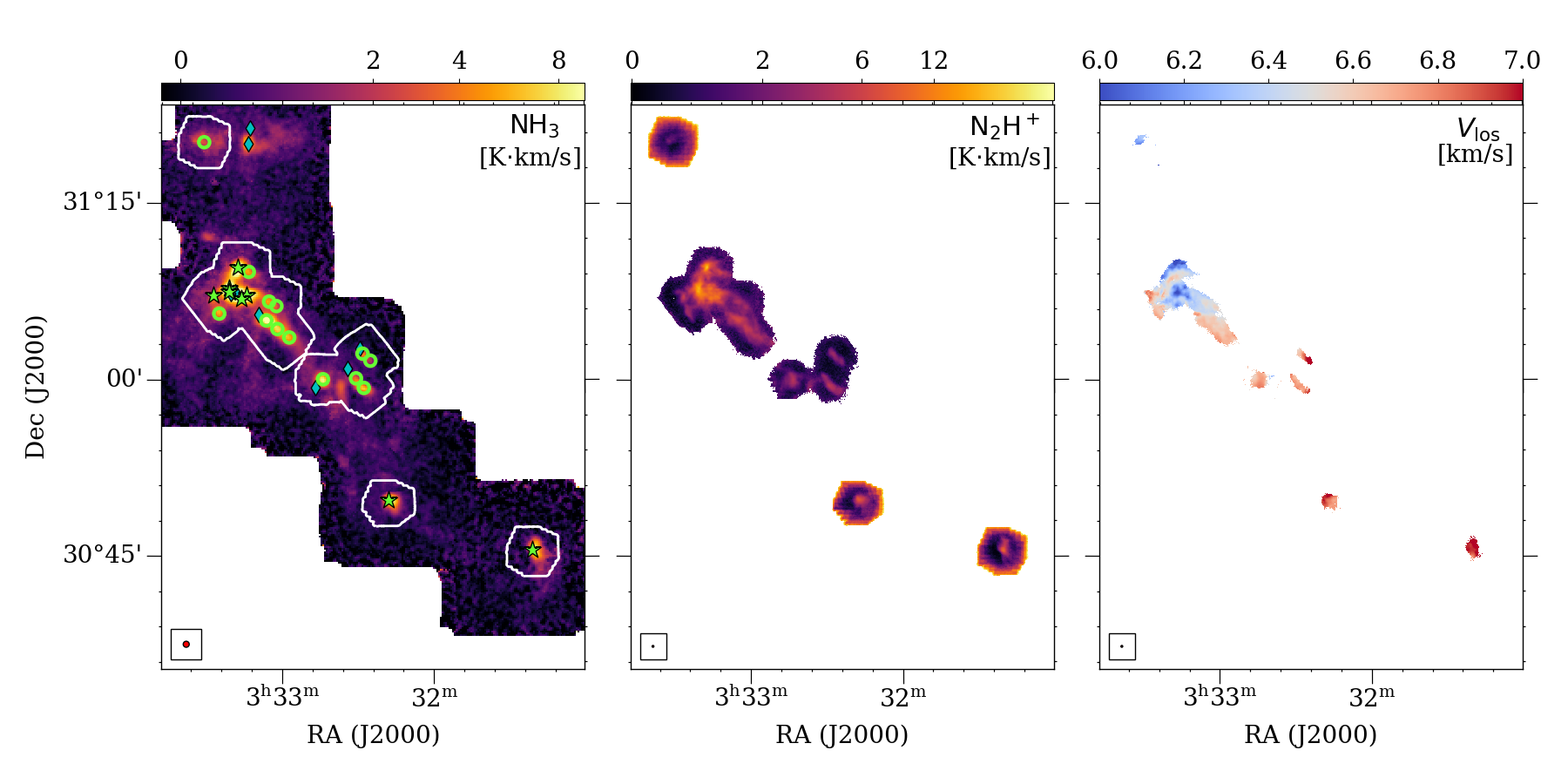}
    \caption{DiSCo observations of B1. {\it Left:} selected Class 0 protostellar ({\it stars}) and starless ({\it open circles}) targets of the DiSCo survey overplotted on the NH$_3$ moment~0 map (in K\,\kms) from GAS \citep{GAS_2017}, with white contours showing the coverage of DiSCo observations. {\it Middle}: DiSCo \NtwoH{} moment~0 map of B1 in K\,\kms. {\it Right:} the fitted line-of-sight \NtwoH{} central velocity in \kms from DiSCo. The 30\arcsec{} GAS beam and 9\arcsec{} GBT-Argus beam are shown in the lower left corner of the corresponding panels. Also plotted in the left panel are the Class I protostars ({\it blue diamonds}) that are not included in DiSCo. }
    \label{fig:B1result}
\end{figure*}

\begin{figure*}
	\includegraphics[width=\textwidth]{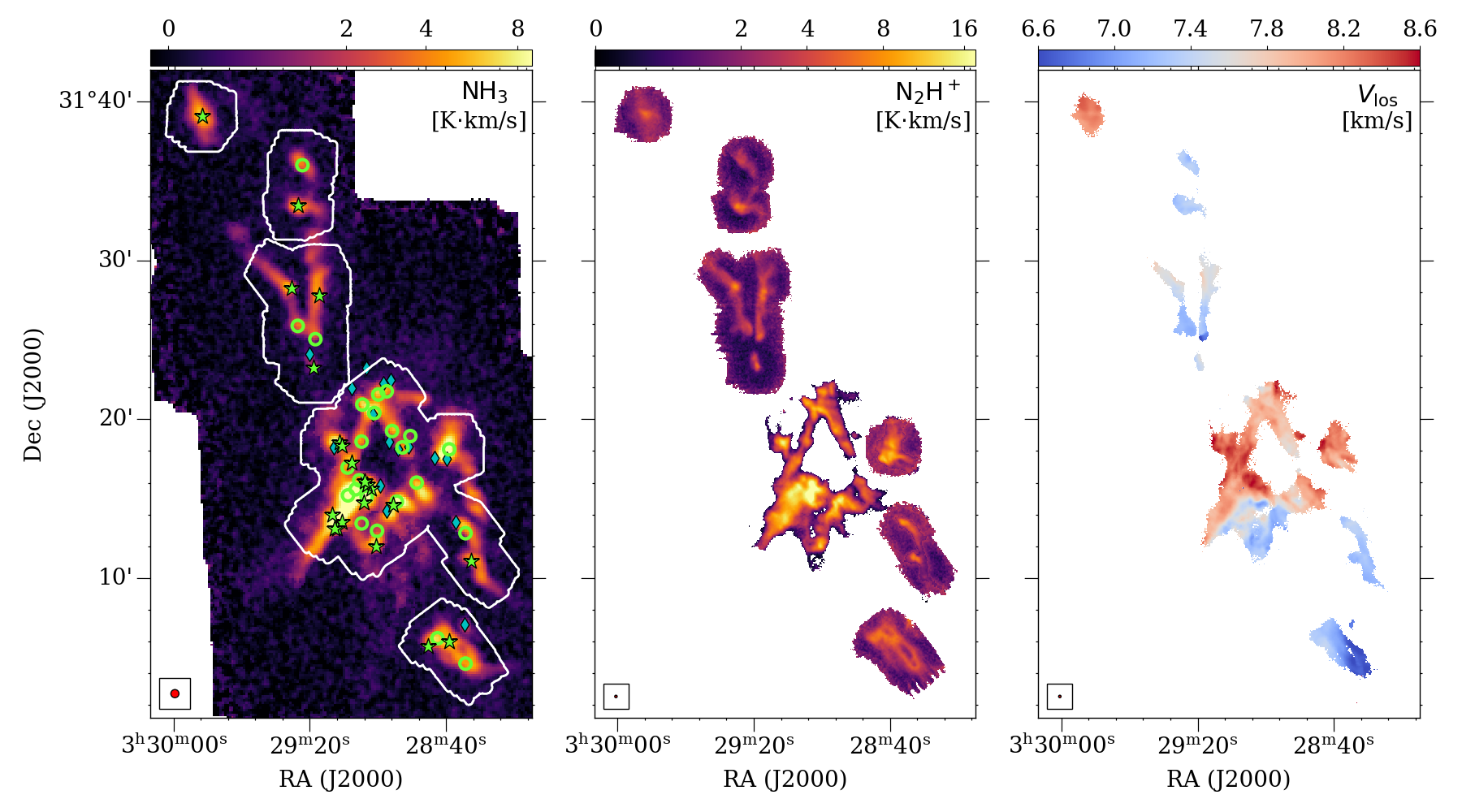}
    \caption{Similar to Fig.~\ref{fig:B1result}, but for NGC\,1333.}
    \label{fig:N1333result}
\end{figure*}

The observations toward B1 and NGC\,1333 of Perseus are summarized in Fig.~\ref{fig:B1result} and \ref{fig:N1333result}. 
Zoom-in velocity maps of individual cores are shown in Fig.~\ref{fig:B1core}-\ref{fig:SVS13} (see Sec.~\ref{sec:core} below). 
\revt{
Here, we use the protostar names from previous studies \citep[e.g.,][]{Tobin_VANDAM_2016} to represent the corresponding protostellar cores (mostly {\it Per-emb-(number)} with some exceptions), and label starless cores as {\it (shortened region name)-(number)} (e.g., B1-2, N1333-5).
}
Tables~\ref{tab:B1}$-$\ref{tab:N1333M} list the measured and derived DiSCo core properties for B1 and NGC\,1333 (single and multi-peak cores), respectively.
\revt{
Note that there are some offsets between the {\it Herschel} column density peak and the DiSCo \NtwoH{} integrated emission peak, but these are mostly insignificant. This could be due to the distribution of \NtwoH{} being not spherically symmetric, or simply caused by resolution effects, because the {\it Herschel} column density resolution is $\sim 4\times$ worse than GBT-{\it Argus} (36.3\arcsec{} vs.~9\arcsec). Some more discussions on peak offset can be found in \citet[][Sec. 4.2]{GBTcore19}.
}

\begin{figure*}
	\includegraphics[width=0.95\textwidth]{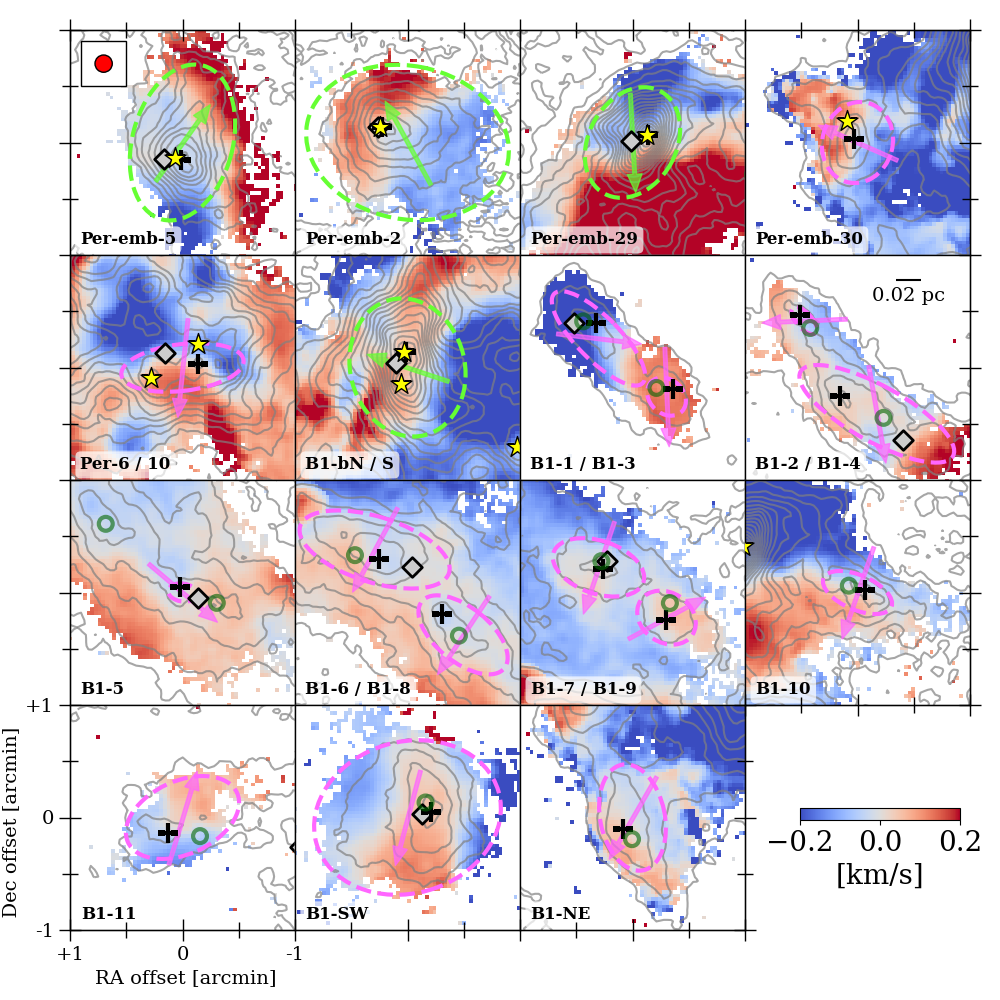}
    \caption{\revt{DiSCo \NtwoH{} maps} for cores in the B1 region. Identified core boundaries ({\it dashed ellipses}) with mean velocity gradient direction ({\it arrows}) overplotted on normalized line-of-sight velocity maps ($\Delta v_{\rm los} \equiv v_{\rm los} - \overline{v_{\rm los}}|_{\rm core}$ in \kms) \revt{and integrated \NtwoH{} intensity ({\it grey contours}; in the level of 1\,K$\cdot$\kms)}. Green ellipses/arrows represent fitting results using {\it Herschel} column density contours, while purple ellipses/arrows showing \revt{cores identified using \NtwoH{} emission}. Yellow stars mark the location of Class 0 protostars, while green circles represent dense cores \revt{(as \NtwoH{} peaks)} identified in previous literature (see Sec.~\ref{sec:target} for more details). Local maxima of {\it Herschel} column density ({\it gray diamonds}) and DiSCo integrated \NtwoH{} emission ({\it cross}) are also marked. The beam is shown in the upper left corner, \revt{and a physical scalebar of 0.02\,pc is shown in the B1-2/B1-4 panel. The coordinates of protostars and \NtwoH{} peaks associated with starless cores are given in Table~\ref{tab:B1}.} }
    \label{fig:B1core}
\end{figure*}

\begin{figure*}
\vspace{-.1in}
	\includegraphics[width=0.89\textwidth]{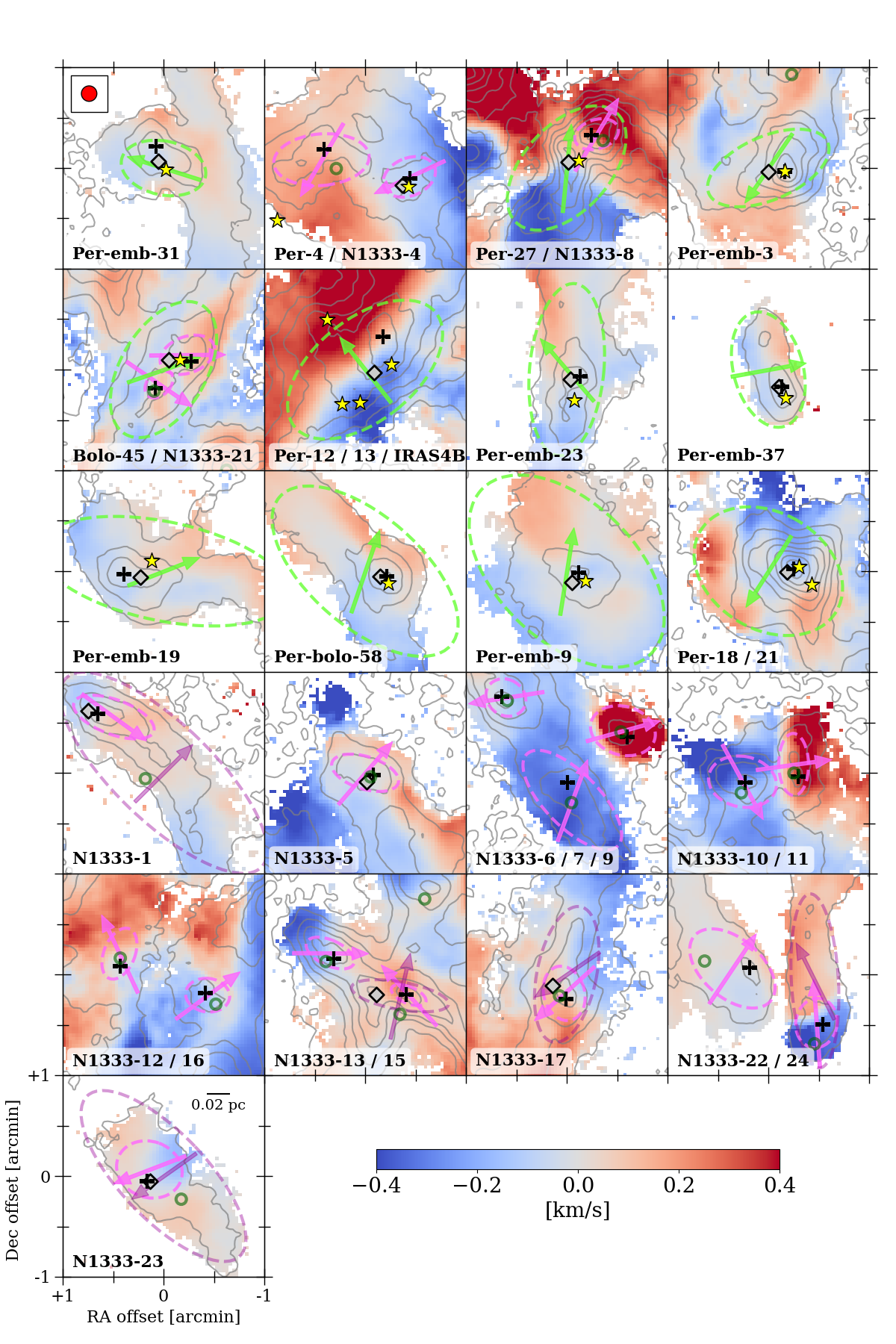}
    \caption{Similar to Fig.~\ref{fig:B1core} but for NGC\,1333 (except core N1333-3 and those in the SVS\,13 region). \revt{A physical scalebar of 0.02\,pc is shown in the N1333-23 panel. The coordinates of protostars and \NtwoH{} peaks associated with starless cores are given in Table~\ref{tab:N1333S}. 
    }}
    \label{fig:N1333core}
\end{figure*}

\begin{figure*}
	\includegraphics[width=\columnwidth]{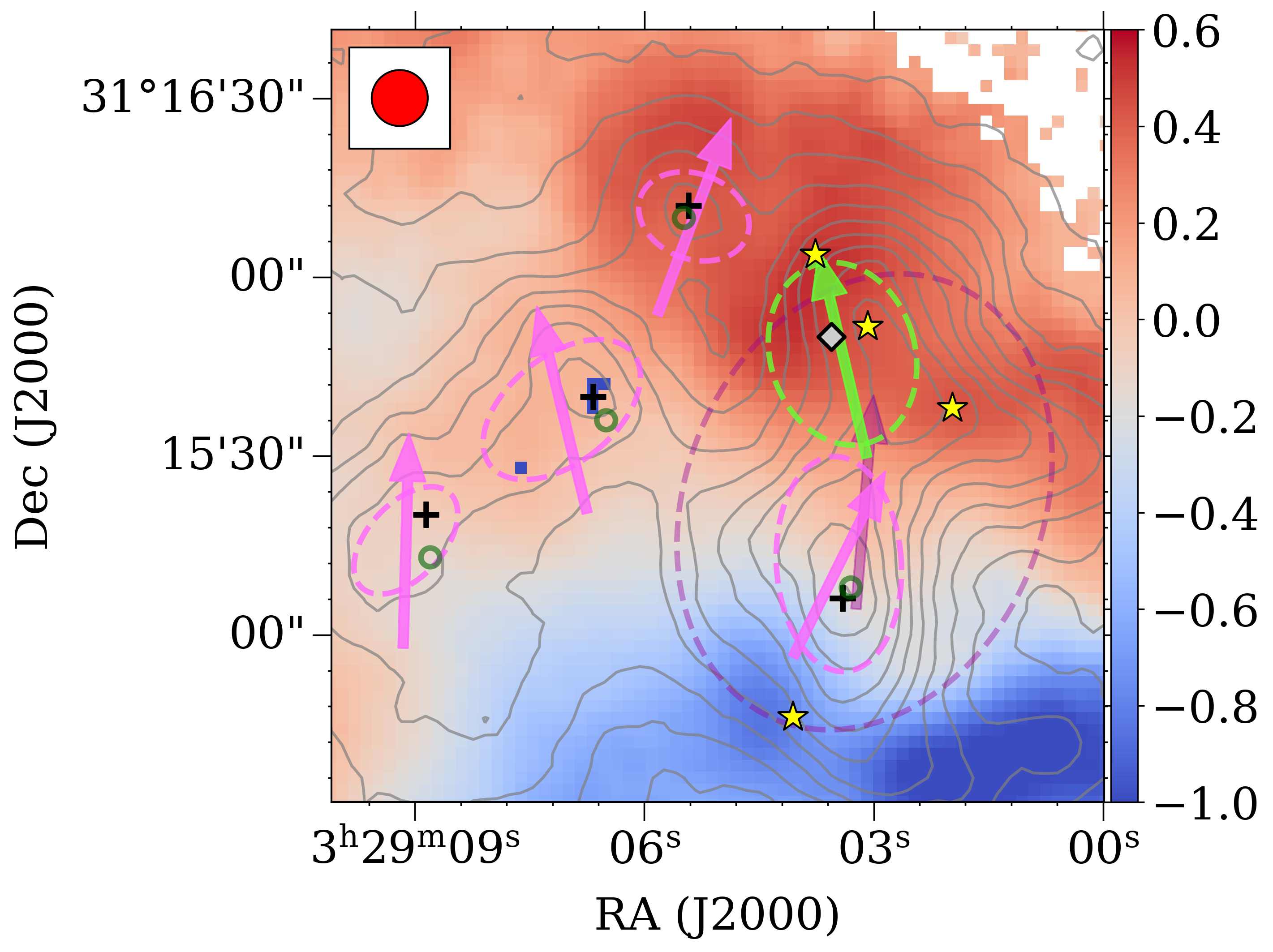}
	\includegraphics[width=\columnwidth]{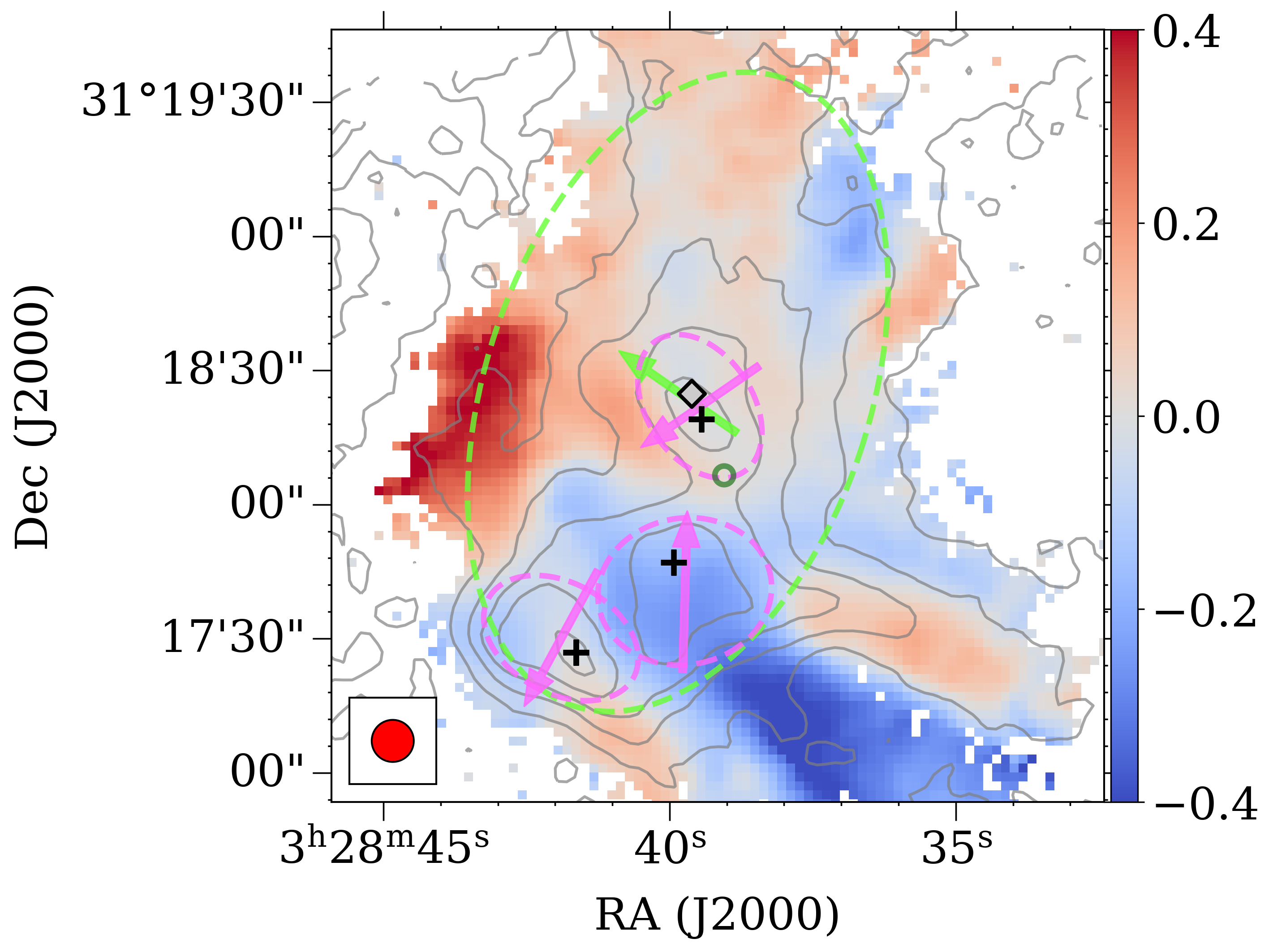}
    \caption{Similar to Fig.~\ref{fig:B1core} but for the SVS13 region ({\it left}) and a multi-peak starless core NGC\,1333-3 ({\it right}).}
    \label{fig:SVS13}
\end{figure*}

\begin{landscape}
\begin{table}
\centering
  \caption{Summary of the targets in the B1 region.}
  \label{tab:B1}
\begin{threeparttable}
    \begin{tabular}{lccccccccccccc}
    \hline
\multirow{2}{*}{target} & \multirow{2}{*}{RA$^\dagger$} & \multirow{2}{*}{Dec$^\dagger$} & IRAM & $\log (N_{\rm H_2, peak}$ & peak $I_{\rm N_2 H^+}$ & BC & aspect & radius & fit & $\overline{v_{\rm los}}$ & $\overline{\nabla v}$ & $\theta_{\overline{\nabla v}}$$^\ddagger$ & $J/10^{-3}$ \\
& & & Src \#$^\blacklozenge$ & $/{\rm cm}^{-2})$ & [K$\cdot$\kms] & level$^*$ & ratio & [pc] & PA [$^\circ$] & [\kms] & [\kms pc$^{-1}$] & [$^\circ$] & [pc\,$\cdot$\,\kms] \\
\hline
Per-emb-5 & 03:31:20.94 & 30:45:30.27 & 90 & 22.45 & \ 9.44 & 22.20$^*$ & 0.63 & 0.045 & -15.1 & 6.99 & 3.1 $\pm$ 2.8 & \ \ -40 $\pm$ 49 & 6.47 \\
Per-emb-2 & 03:32:17.93 & 30:49:47.73 & 86 & 22.76 & \ 8.40 & 22.15$^*$ & 0.75 & 0.071 & \ 79.0 & 6.84 & 2.2 $\pm$ 2.9 & \ \ \ 33 $\pm$ 53 & 11.09~~ \\
Per-emb-6 & 03:33:14.40 & 31:07:10.72 & \multirow{2}{*}{ 73 } & \multirow{2}{*}{ 22.82 } & \multirow{2}{*}{ 11.36 } & \multirow{2}{*}{ 9.66 } & \multirow{2}{*}{ 0.38 } & \multirow{2}{*}{ 0.027 } & \multirow{2}{*}{ -83.2 } & \multirow{2}{*}{ 6.28 } & \multirow{2}{*}{ 6.9 $\pm$ 4.4 } & \multirow{2}{*}{ \ 173 $\pm$ 38 } & \multirow{2}{*}{ 5.03 } \\
Per-emb-10 & 03:33:16.42 & 31:06:52.06 & & & & & & & & & & \\
Per-emb-29 & 03:33:17.88 & 31:09:31.82 & 72 & 22.85 & 20.09 & 22.65$^*$ & 0.75 & 0.039 & -29.0 & 6.18 & 6.3 $\pm$ 3.4 & -176 $\pm$ 36 & 9.48 \\
B1-bN & 03:33:21.21 & 31:07:43.67 & \multirow{2}{*}{ 71 } & \multirow{2}{*}{ 23.10 } & \multirow{2}{*}{ 18.42 } & \multirow{2}{*}{ 22.75$^*$ } & \multirow{2}{*}{ 0.82 } & \multirow{2}{*}{ 0.054 } & \multirow{2}{*}{ \ 14.5 } & \multirow{2}{*}{ 6.46 } & \multirow{2}{*}{ 2.7 $\pm$ 4.7 } & \multirow{2}{*}{ \ \ \ 74 $\pm$ 50 } & \multirow{2}{*}{7.83} \\
B1-bS & 03:33:21.36 & 31:07:26.37 & & & & & & & & & & \\
Per-emb-30 & 03:33:27.30 & 31:07:10.16 & 68 & 22.39 & \ 5.62 & 3.12 & 0.86 & 0.029 & -15.5 & 6.69 & 5.7 $\pm$ 3.9 & \ \ \ 69 $\pm$ 37 & 4.81 \\
\hline
B1-1 & 03:32:24.38 & 31:01:36.12 & & 22.09 & \ 2.92 & 2.42 & 0.79 & 0.015 & \ 47.0 & 7.02 & 1.0 $\pm$ 0.9 & -177 $\pm$ 57 & 0.22 \\
B1-3 & 03:32:27.65 & 31:02:12.12 & 84 & 22.11 & \ 3.48 & 2.38 & 0.34 & 0.029 & \ 44.6 & 6.77 & 5.9 $\pm$ 3.2 & \ \ -96 $\pm$ 24 & 4.87\\
B1-2 & 03:32:29.52 & 30:59:30.12 & 85 & 22.38 & \ 5.20 & 3.20 & 0.32 & 0.039 & \ 61.0 & 6.76 & 1.4 $\pm$ 2.3 & -168 $\pm$ 67 & 2.22\\
B1-4 & 03:32:31.22 & 31:00:14.04 & &  22.38 & \ 3.37 & 3.17 & 0.45 & 0.009 & \ 41.8 & 6.70 & 4.1 $\pm$ 1.3 & \ \ \ 92 $\pm$ 20 & 0.34\\
B1-5 & 03:32:58.85 & 31:03:43.92 & 77 & 22.53 & \ 4.40 & 4.20 & 0.62 & 0.009 & \ 81.5 & 6.63 & 2.5 $\pm$ 1.1 & -126 $\pm$ 32 & 0.22 \\
B1-6 & 03:33:02.74 & 31:04:30.00 & & 22.56 & \ 5.49 & 4.49 & 0.51 & 0.030 & \ 50.9 & 6.60 & 0.9 $\pm$ 1.4 & \ 142 $\pm$ 66 & 0.76 \\
B1-8 & 03:33:05.38 & 31:04:59.88 & 75 & 22.56 & \ 6.83 & 4.53 & 0.42 & 0.040 & \ 72.5 & 6.61 & 1.2 $\pm$ 2.2 & \ 148 $\pm$ 60 & 1.98\\
B1-7 & 03:33:02.57 & 31:06:06.12 & & 22.56 & \ 3.97 & 2.97 & 0.80 & 0.022 & \ 56.1 & 6.53 & 2.4 $\pm$ 2.0 & \ \ -65 $\pm$ 49 & 1.17 \\
B1-9 & 03:33:05.23 & 31:06:33.84 & & 22.41 & \ 4.25 & 3.05 & 0.56 & 0.028 & \ 71.6 & 6.50 & 1.7 $\pm$ 1.9 & \ 159 $\pm$ 52 & 1.36 \\
B1-10 & 03:33:12.70 & 31:09:07.92 & & 22.85 & \ 6.43 & 5.13 & 0.47 & 0.020 & \ 66.9 & 6.41 & 5.1 $\pm$ 1.9 & \ 158 $\pm$ 22 & 2.03 \\
B1-11 & 03:33:32.57 & 31:20:12.84 & 67 & 22.24 & \ 2.48 & 1.38 & 0.61 & 0.037 & -65.7 & 6.27 & 2.4 $\pm$ 2.2 & \ \ -20 $\pm$ 49 & 3.38 \\
B1-SW & 03:32:43.68 & 30:59:57.12 & 79 & 22.46 & \ 4.36 & 0.66 & 0.80 & 0.067 & -74.1 & 6.79 & 0.8 $\pm$ 2.9 & \ 163 $\pm$ 69 & 3.71 \\
B1-NE & 03:33:25.46 & 31:05:42.00 & 69 & $-^\S$ & \ 3.83 & 2.63 & 0.60 & 0.033 & \ 11.4 & 6.64 & 1.7 $\pm$ 3.0 & \ 146 $\pm$ 46 & 1.83\\
\hline
\end{tabular}
 \begin{tablenotes}
      \small
      \item $^\dagger$Coordinates for protostellar cores (upper half in the table) are the protostar coordinates from the VANDAM survey \citep{Tobin_VANDAM_2016}. Starless core (lower half in the table) coordinates are the observed peak \NtwoH{} locations.
      \item $^\blacklozenge$The corresponding \NtwoH{} source number in \cite{Kirk_Perseus_2007} and \cite{Johnstone_NH3_2010}.
      \item $^*$Core boundary contours (BC) are defined using either {\it Herschel} H$_2$ column densities (marked with $^*$) or observed \NtwoH{} integrated intensities; see Sec.~\ref{sec:core}.
      \item $^\ddagger$Here, the uncertainty of $\theta_{\overline{\nabla v}}$ is estimated as the circular mean of $\theta_{\nabla v} - \theta_{\overline{\nabla v}}$. See Sec.~\ref{sec:rotation} for more discussions on $\sigma_{\measuredangle(\nabla v, \overline{\nabla v})}$.
      \item $^\S$B1-NE is too close to the B1-b region and thus is not resolved in {\it Herschel}.
    \end{tablenotes}
  \end{threeparttable}
\end{table}
\end{landscape}

\begin{landscape}
\begin{table}
 \centering
  \caption{Similar to Table~\ref{tab:B1}, but for single-peak targets in the NGC\,1333 region.}
  \label{tab:N1333S}
\begin{threeparttable}
    \begin{tabular}{lccccccccccccc}
    \hline
\multirow{2}{*}{target} & \multirow{2}{*}{RA} & \multirow{2}{*}{Dec} & IRAM & $\log (N_{\rm H_2, peak}$ & peak $I_{\rm N_2 H^+}$ & BC & aspect & radius & fit & $\overline{v_{\rm los}}$ & $\overline{\nabla v}$ & $\theta_{\overline{\nabla v}}$ & $J/10^{-3}$ \\
& & & Src \# & $/{\rm cm}^{-2})$ & [K$\cdot$\kms] & level & ratio & [pc] & PA [$^\circ$] & [\kms] & [\kms pc$^{-1}$] & [$^\circ$] & [pc\,$\cdot$\,\kms] \\
\hline
Per-emb-31 & 03:28:32.55 & 31:11:05.15 & 128 & 22.16 & 7.76 & 22.06$^*$ & 0.63 & 0.034 & 78.3 & 7.21 & 2.1 $\pm$ 2.5 & \ \ 75 $\pm$ 49 & 2.34 \\
Per-emb-4 & 03:28:39.10 & 31:06:01.80 & & 22.36 & 6.95 & 5.95 & 0.69 & 0.021 & 112.5~~ & 7.05 & 3.5 $\pm$ 2.1 & 112 $\pm$ 14 & 1.58 \\
N1333-4 & 03:28:43.08 & 31:06:24.84 & 121 & 22.36 & 6.94 & 5.94 & 0.54 & 0.031 & -85.3~ & 7.27 & 2.8 $\pm$ 2.2 & 145 $\pm$ 40 & 2.70 \\
Per-emb-3 & 03:29:00.58 & 31:12:00.20 & 115 & 22.40 & 14.40~~ & 22.30$^*$ & 0.49 & 0.036 & -66.6~ & 7.21 & 2.0 $\pm$ 2.8 & 141 $\pm$ 71 & 2.51 \\
Per-emb-23 & 03:29:17.21 & 31:27:46.31 & 100 & 22.36 & 7.17 & 22.21$^*$ & 0.43 & 0.045 & \ -5.5 & 7.57 & 2.4 $\pm$ 2.5 & \ \ 45 $\pm$ 58 & 4.95 \\
Per-emb-37 & 03:29:18.97 & 31:23:14.30 & 98 & 22.30 & 6.25 & 22.15$^*$ & 0.59 & 0.042 & 15.0 & 7.49 & 4.2 $\pm$ 2.9 & -81 $\pm$ 33 & 7.44 \\
Per-emb-19 & 03:29:23.50 & 31:33:29.17 & 97 & 22.38 & 7.59 & 21.93$^*$ & 0.39 & 0.072 & 78.4 & 7.32 & 1.7 $\pm$ 2.0 & -72 $\pm$ 61 & 8.73 \\
Per-bolo-58 & 03:29:25.46 & 31:28:14.88 & 96 & 22.29 & 8.44 & 21.89$^*$ & 0.49 & 0.068 & 49.1 & 7.54 & 2.3 $\pm$ 2.9 & -22 $\pm$ 57 & 10.4 \\
Per-emb-9 & 03:29:51.83 & 31:39:05.90 & 95 & 22.32 & 6.26 & 21.87$^*$ & 0.59 & 0.080 & 45.9 & 8.21 & 1.7 $\pm$ 2.2 & -11 $\pm$ 61 & 10.8 \\
\hline
N1333-1 & \multirow{2}{*}{ 03:28:36.46 } & \multirow{2}{*}{ 31:13:28.92 }  & & \multirow{2}{*}{ 22.21 } & \multirow{2}{*}{ 9.35 } & 6.15 & 0.41 & 0.024 & 70.0 & 7.37 & 3.0 $\pm$ 2.2 & -122 $\pm$ 49~ & 1.70 \\
(1-big) & & & & & & 2.75 & 0.36 & 0.072 & 46.1 & 7.33 & 1.2 $\pm$ 2.2 & -50 $\pm$ 66 & 6.34 \\
N1333-5 & 03:28:48.43 & 31:16:01.92 & & 22.29 & 7.41 & 6.01 & 0.47 & 0.021 & 72.7 & 8.10 & 2.7 $\pm$ 2.0 & -45 $\pm$ 32 & 1.16\\
N1333-6 & 03:28:50.14 & 31:18:54.00 & & 22.58 & 4.77 & 2.97 & 0.74 & 0.025 & 75.0 & 8.40 & 8.5 $\pm$ 15.3 & -77 $\pm$ 54 & 5.12 \\
N1333-7 & 03:28:52.94 & 31:18:25.92 & & 22.50 & 7.94 & 4.94 & 0.42 & 0.037 & 44.6 & 7.68 & 1.4 $\pm$ 2.8 & -23 $\pm$ 69 & 1.97\\
N1333-9 & 03:28:56.06 & 31:19:18.12 & & 22.45 & 10.76~~ & 8.36 & 0.86 & 0.017 & 57.1 & 7.90 & 4.2 $\pm$ 2.3 & \ \ 98 $\pm$ 21 & 1.28\\
N1333-10 & 03:28:57.17 & 31:21:43.92 & & 22.78 & 8.71 & 6.91 & 0.48 & 0.019 & \ \ 0.9 & 8.35 & 20.7 $\pm$ 10.7 & -83 $\pm$ 20 & 7.27 \\
N1333-11 & 03:28:59.66 & 31:21:39.96 & & 22.78 & 10.75~~ & 6.95 & 0.74 & 0.026 & 78.9 & 7.80 & 6.2 $\pm$ 4.6 & -148 $\pm$ 36~~ & 4.24\\
N1333-12 & 03:29:00.74 & 31:13:05.88 & & 22.99 & 6.98 & 5.78 & 0.76 & 0.017 & 83.2 & 7.32 & 6.8 $\pm$ 3.2 & -58 $\pm$ 32 & 1.99 \\
N1333-16 & 03:29:04.80 & 31:13:22.08 & & 23.20 & 5.86 & 5.06 & 0.58 & 0.018 & -21.9 & 7.51 & 7.6 $\pm$ 4.4 & \ \ 29 $\pm$ 32 & 2.44\\
N1333-13 & \multirow{2}{*}{ 03:29:00.74 } & \multirow{2}{*}{ 31:20:35.88 } & \multirow{2}{*}{ 113 } & \multirow{2}{*}{ 22.89 } & \multirow{2}{*}{ 12.75~~ } & 11.75~~ & 0.45 & 0.011 & 67.2 & 8.05 & 5.1 $\pm$ 2.6 & \ \ 47 $\pm$ 28 & 0.61 \\
(13-big) & & & & & & 10.55~~ & 0.28 & 0.023 & 79.9 & 8.00 & 2.2 $\pm$ 3.3 & -15 $\pm$ 71 & 1.15 \\
N1333-14 & 03:29:03.41 & 31:15:06.12 & & 23.20 & 26.09~~ & 21.69~~ & 0.58 & 0.020 & \ \ \ 3.4 & 7.96 & 13.3 $\pm$ 4.6 & -30 $\pm$ 15 & 5.15 \\
N1333-18 & 03:29:05.42 & 31:16:12.00 & & 22.98 & 12.64~~ & 11.24~~ & 0.75 & 0.012 & 69.4 & 8.48 & 2.5 $\pm$ 1.5 & -24 $\pm$ 47 & 0.35 \\
N1333-19 & 03:29:06.67 & 31:15:39.96 & 110 & 23.18 & 18.98~~ & 16.58~~ & 0.59 & 0.017 & -51.8~ & 8.14 & 3.3 $\pm$ 2.6 & \ \ 16 $\pm$ 43 & 0.93 \\
N1333-20 & 03:29:08.86 & 31:15:20.16 & 107 & 22.98 & 17.25~~ & 16.65~~ & 0.58 & 0.012 & -43.0~ & 7.99 & 4.1 $\pm$ 2.8 & \ \ -2 $\pm$ 33 & 0.57 \\
N1333-15 & 03:29:04.18 & 31:20:57.84 & & 22.89 & 11.47~~ & 9.47 & 0.49 & 0.016 & 65.1 & 7.80 & 8.4 $\pm$ 2.5 & -91 $\pm$ 15 & 2.19 \\
N1333-17 & \multirow{2}{*}{ 03:29:04.49 } & \multirow{2}{*}{ 31:18:33.84 } & & \multirow{2}{*}{ 22.53 } & \multirow{2}{*}{ 8.47 } & 6.67 & 0.74 & 0.021 & 15.0 & 8.21 & 7.6 $\pm$ 3.8 & 128 $\pm$ 28 & 3.35 \\
(17-big) & & & & & & 5.07 & 0.43 & 0.040 & 11.6 & 8.15 & 4.3 $\pm$ 3.6 & 121 $\pm$ 46 & 6.81 \\
N1333-22 & \multirow{2}{*}{ 03:29:18.02 } & \multirow{2}{*}{ 31:25:14.88 } & \multirow{2}{*}{ 99 } & \multirow{2}{*}{ 22.33 } & \multirow{2}{*}{ 5.83 } & 4.03 & 0.83 & 0.020 & \ \ -9.2 & 6.94 & 13.7 $\pm$ 15.1 & \ \ \ 5 $\pm$ 46 & 5.39 \\
(22-big) & & & & & & 3.03 & 0.27 & 0.040 & \ \ \ 3.2 & 7.14 & 5.0 $\pm$ 10.8 & \ \ 30 $\pm$ 64 & 8.24 \\
N1333-24 & 03:29:21.48 & 31:25:49.08 & & 22.36 & 4.43 & 2.83 & 0.59 & 0.034 & 49.4 & 7.14 & 1.4 $\pm$ 2.1 & -38 $\pm$ 67 & 1.62 \\
N1333-23 & \multirow{2}{*}{ 03:29:23.86 } & \multirow{2}{*}{ 31:36:11.16 } & & \multirow{2}{*}{ 22.12 } & \multirow{2}{*}{ 3.99 } & 2.99 & 0.84 & 0.027 & 70.4 & 7.28 & 3.4 $\pm$ 2.4 & 108 $\pm$ 37 & 2.50 \\
(23-big) & & & & & & 1.79 & 0.39 & 0.061 & 43.6 & 7.31 & 1.6 $\pm$ 2.5 & 121 $\pm$ 58 & 6.12 \\
\hline
\end{tabular}
%  \tablefoot{
%    \tablefoottext{a}{B1-NE is too close to the B1-b region and thus is not resolved in {\it Herschel}.}
%    \tablefoottext{b}{In the units that $v$ in km\,s$^{-1}$ and $\ell$ in pc.}
%    \tablefoottext{c}{Using $\sin{\cal S}_{\rm corr}$ here.}
%  }
\end{threeparttable}
\end{table}
\end{landscape}

\begin{landscape}
\begin{table}
 \centering
  \caption{Similar to Table~\ref{tab:N1333S}, but for multi-peak targets in NGC\,1333. 
  \revt{For each multi-peak core (separated by horizontal lines), we listed the identified \NtwoH{} peaks within the core (protostellar sources first, then starless sources) and the corresponding fitting results. We prioritize our core identification routine to have as few \NtwoH{} peaks within a core as possible, but we also provide the measurement using core boundaries identified with {\it Herschel} column density. For example, Bolo-45 and N1333-21 each has a \NtwoH-defined core boundary, but they belong to the same dense core when {\it Herschel} column density is used to define the core boundary.
  On the other hand, some sources are too clustered to have separate core boundaries, and we hence provide measurements inside the core boundaries that include multiple peaks (e.g.,~the SVS\,13 region). }
  }
  \label{tab:N1333M}
\begin{threeparttable}
    \begin{tabular}{lccccccccccccc}
    \hline
\multirow{2}{*}{target} & \multirow{2}{*}{RA} & \multirow{2}{*}{Dec} & IRAM & $\log (N_{\rm H_2, peak}$ & peak $I_{\rm N_2 H^+}$ & BC & aspect & radius & fit & $\overline{v_{\rm los}}$ & $\overline{\nabla v}$ & $\theta_{\overline{\nabla v}}$ & $J/10^{-3}$ \\
& & & Src \# & $/{\rm cm}^{-2})$ & [K$\cdot$\kms] & level & ratio & [pc] & PA [$^\circ$] & [\kms] & [\kms pc$^{-1}$] & [$^\circ$] & [pc\,$\cdot$\,\kms] \\
\hline
Per-emb-27 & \multirow{2}{*}{03:28:55.57} & \multirow{2}{*}{31:14:37.03} & \multirow{3}{*}{118} & \multirow{3}{*}{22.70} & \multirow{3}{*}{18.21} & \multirow{2}{*}{22.45$^*$} & \multirow{2}{*}{0.58} & \multirow{2}{*}{0.047} & \multirow{2}{*}{-42.4} & \multirow{2}{*}{7.50} & \multirow{2}{*}{7.9 $\pm$ 10.0} & \multirow{2}{*}{\ \ \ -7 $\pm$ 47} & \multirow{2}{*}{17.45~~} \\
(+N1333-8) & \\
N1333-8 & 03:28:54.98 & 31:14:52.08 & & & & 15.21~ & 0.65 & 0.016 & -72.8 & 7.81 & 14.3 $\pm$ 8.3~~ & \ \ -35 $\pm$ 17 & 3.73 \\
\hline
Bolo-45 & 03:29:07.70 & 31:17:16.80 & \multirow{3}{*}{109} & \multirow{3}{*}{22.70} & \ \ 9.30 & 8.10 & 0.64 & 0.017 & -69.5 & 8.42 & 4.3 $\pm$ 2.5 & \ \ -90 $\pm$ 25 & 1.18 \\
N1333-21 & 03:29:08.92 & 31:16:57.94 & & & \ \ 9.10 & 8.10 & 0.70 & 0.012 & -44.2 & 8.38 & 3.9 $\pm$ 4.5 & -120 $\pm$ 50 & 0.57 \\
(both) & $-$ & $-$ & & & \ \ 9.30 & 22.50$^*$ & 0.57 & 0.050 & -30.8 & 8.34 & 1.9 $\pm$ 3.8 & \ \ -74 $\pm$ 68 & 4.83 \\
\hline
Per-emb-12 & 03:29:10.54 & 31:13:30.93 & 106 & \multirow{3}{*}{ 23.20 } & \multirow{3}{*}{ 15.92 } & \multirow{3}{*}{ 22.80$^*$ } & \multirow{3}{*}{ 0.56 } & \multirow{3}{*}{ 0.057 } & \multirow{3}{*}{ -51.05 } & \multirow{3}{*}{ 7.62 } & \multirow{3}{*}{ 9.1 $\pm$ 5.4 } & \multirow{3}{*}{ \ \ \ 43 $\pm$ 32 } & \multirow{3}{*}{ 29.63~~ } \\
Per-emb-13 & 03:29:12.02 & 31:13:08.03 & 103 & & & & & & & & & & \\
IRAS4B' & 03:29:12.84 & 31:13:06.90 & & & & & & & & & & & \\
\hline
SVS13C & 03:29:01.97 & 31:15:38.05 & \multirow{5}{*}{ 112 } & \multirow{5}{*}{ 22.91 } & \multirow{5}{*}{ 26.09 } & \multirow{4}{*}{ 14.89~ } & \multirow{4}{*}{ 0.76 } & \multirow{4}{*}{ 0.052 } & \multirow{4}{*}{ -22.0 } & \multirow{4}{*}{ 8.08 } & \multirow{4}{*}{ 10.0 $\pm$ 7.2 } & \multirow{4}{*}{ \ \ \ -6 $\pm$ 35 } & \multirow{4}{*}{ 27.14~~ } \\
SVS13B & 03:29:03.08 & 31:15:51.74 & & & & & & & & & & & \\
Per-emb-44 & 03:29:03.76 & 31:16:03.81 & & & & & & & & &  & \\
Per-emb-15 & 03:29:04.06 & 31:14:46.24 & & & & & & & & & & & \\
SVS13B+C & $-$ & $-$ & & & & 22.86$^*$ & 0.78 & 0.021 & \ \ 16.5 & 8.50 & 6.0 $\pm$ 4.3 & \ \ 15 $\pm$ 41 & 2.51 \\
\hline
N1333-3 & 03:28:39.55 & 31:18:24.84 & \multirow{4}{*}{ 123 } & \multirow{4}{*}{ 22.66 } & 12.64 & 22.31$^*$ & 0.58 & 0.082 & -18.5 & 8.18 & 2.6 $\pm$ 4.5 & \ \ 59 $\pm$ 67 & 17.61~~ \\
(3-A) & 03:28:41.57 & 31:17:26.88 & & & 12.64 & 9.24 & 0.67 & 0.022 & \ \ 61.4 & 8.12 & 2.6 $\pm$ 4.5 & \ 148 $\pm$ 74 & 1.22 \\
(3-B) & 03:28:39.86 & 31:17:47.04 & & & 11.65 & 9.25 & 0.84 & 0.026 & -80.4 & 7.98 & 2.9 $\pm$ 3.1 & \ \ -1.9 $\pm$ 51 & 1.97 \\
(3-C) & 03:28:39.38 & 31:18:19.08 & & & 10.90 & 8.50 & 0.67 & 0.021 & \ \ 33.8 & 8.19 & 2.1 $\pm$ 3.4 & \ 121 $\pm$ 74 & 0.89 \\
\hline
\end{tabular}
%  \tablefoot{
%    \tablefoottext{a}{B1-NE is too close to the B1-b region and thus is not resolved in {\it Herschel}.}
%    \tablefoottext{b}{In the units that $v$ in km\,s$^{-1}$ and $\ell$ in pc.}
%    \tablefoottext{c}{Using $\sin{\cal S}_{\rm corr}$ here.}
%  }
\end{threeparttable}
\end{table}
\end{landscape}

\subsection{Core Identification}
\label{sec:core}

\begin{figure*}
	\includegraphics[width=\columnwidth]{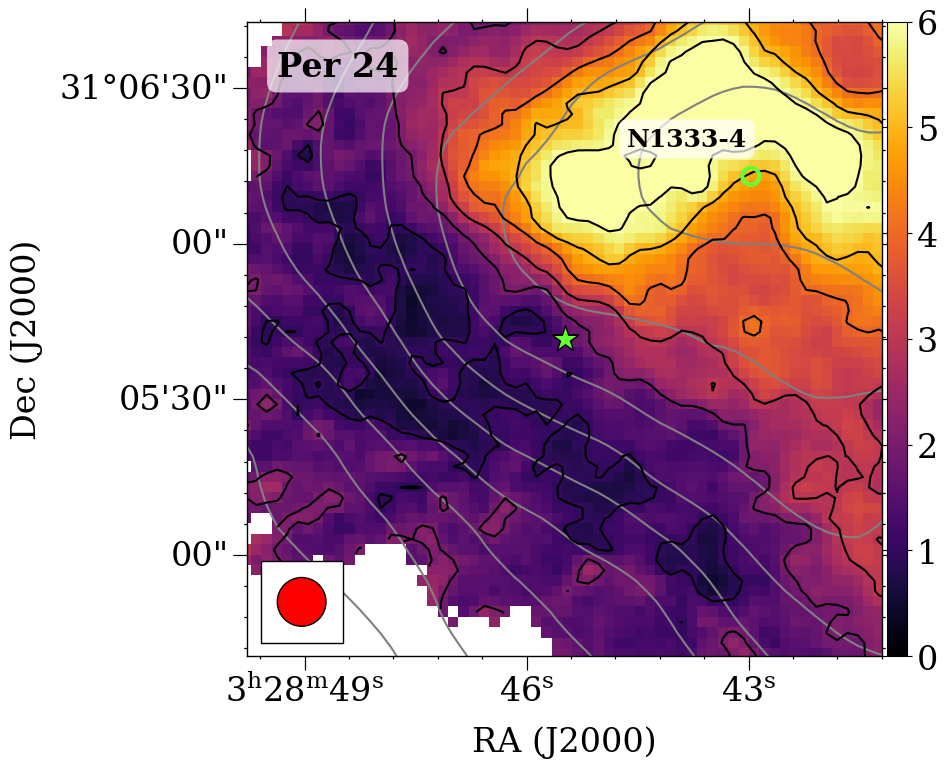}
	\includegraphics[width=\columnwidth]{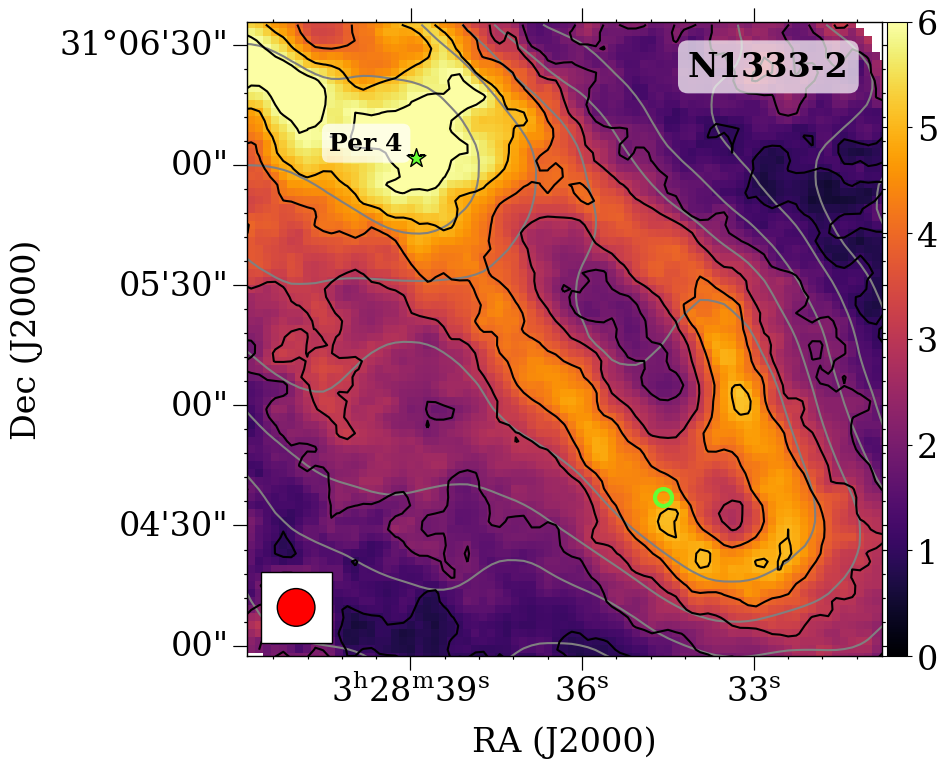}
    \caption{Two previously-defined cores \revt{({\it left:} Class\,0 protostar Per-24, marked by a green star; {\it right}: a starless core, marked by a green circle)} 
    with no core-like structure in integrated \NtwoH{} emission (in K\,\kms, {\it colormap and black contours \revt{in the level of 1\,K$\cdot$\kms}}) from DiSCo. Column density ($N_{\rm H}$ in cm$^{-2}$, \revt{in the level of $\Delta\log (N_{\rm H}/{\rm cm}^{2}) = 0.1$}) contours from {\it Herschel} are plotted as light gray contours.
    \revt{The beam is shown in the bottom left corner in each panel, and other DiSCo targets in the fields of view are also marked.}
    }
    \label{fig:nocore}
\end{figure*}

\begin{figure}
	\includegraphics[width=\columnwidth]{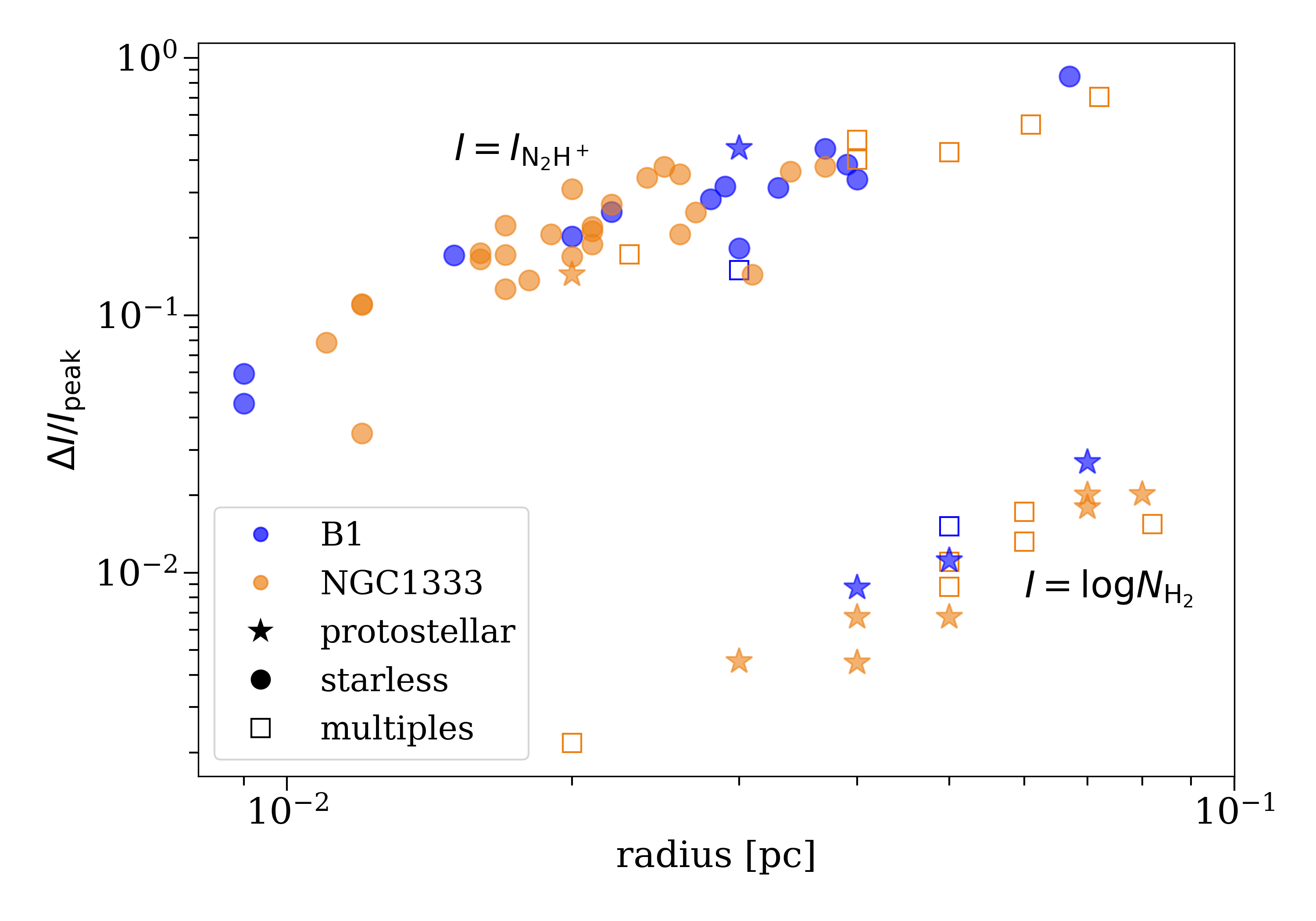}
    \vspace{-.3in}
    \caption{Normalized intensity variation plotted against the size of the corresponding cores, from both B1 ({\it blue symbols}) and NGC\,1333 ({\it yellow symbols}) regions and for both starless ({\it circles}) and protostellar ({\it stars}) cores, single or multiple ({\it open squares}). The values of $\Delta I/I_{\rm peak}$ offsets between protostellar cores measured in {\it Herschel} $\log N_{\rm H_2}$ column densities (lower values of $\Delta I/I_{\rm peak}$) and cores (mostly starless) measured in \NtwoH{} emission.}
    \label{fig:dI}
\end{figure}

\revt{
In general, gas column density is the preferred data for measuring the dense core boundary, because it is fitted from dust thermal radiation and thus is less biased by excitation conditions (e.g., gas density or temperature) than spectral lines. However, since dense cores commonly consist of cold, dense gas that favors molecular lines like \NtwoH, NH$_3$, or C$^{18}$O, these tracers are often considered good core material tracers (see e.g., GAS, CLASSy).\footnote{\revt{While we do not include a direct comparison between column density and \NtwoH{} emission in this work, some insight can be inferred from Figures~\ref{fig:B1core} and \ref{fig:N1333core} by comparing the elliptic core boundaries fitted from column density and the \NtwoH{} integrated emission contours. The column density-defined core boundaries and the brightest regions in \NtwoH{} mostly overlap, suggesting that \NtwoH{} is a good substitute of column density to trace core material. }} 
}

Following our pilot study \citep{GBTcore19}, we consider either the {\it Herschel} column density or \NtwoH{} integrated emission (moment 0) contours as the core boundaries. For protostellar cores, {\it Herschel} column density is used unless there is no closed contour available (due to resolution or clustered environment); in such cases, we considered \NtwoH{} integrated emission instead. On the other hand, all starless cores are defined using \NtwoH{} integrated emission, \revt{because very few of them are resolved in {\it Herschel} and we would like to keep our analysis as consistent as possible.}

Instead of finding a background value by fitting a Gaussian shape to the intensity (as was done in \citealt{GBTcore19}), we adopted a similar approach as considered in the \texttt{GRID-core} algorithm \citep{GO11, CO14, CO15, GO15}, and identify core boundaries as the biggest closed contours that only include one local maximum (``boundary contours''; see column ``BC~level'' in Tables~\ref{tab:B1}$-$\ref{tab:N1333M}). These contours were drawn from the peak value of each core (see Tables~\ref{tab:B1}-\ref{tab:N1333M}) with negative stepsize $\Delta\log (N_{\rm H}/{\rm cm}^{-2}) = 0.05$ for {\it Herschel} contours, and $\Delta I_{\rm N_2 H^+} = 0.1$ (B1) and $0.2\,{\rm K}\cdot$\,\kms{} (NGC\,1333) for \NtwoH{} contours. 
\revt{We note that the biggest advantage of defining cores using contours is that this method better preserves the natural shape of the core, while Gaussian fitting implicitly presumes axisymmetry in cores.}

We calculated most of the core properties (including the mean line-of-sight velocity $\overline{v_{\rm los}}$, mean velocity gradient $\overline{\nabla v}$, and the dispersion of the velocity gradient direction $\theta_{\overline{\nabla v}}$) using these core-boundary contours, and fit such contours with ellipses (see Figs.~\ref{fig:B1core}--\ref{fig:SVS13}) to estimate the sizes, aspect ratios, and orientations (column ``fit PA'') of the cores (Tables~\ref{tab:B1}--\ref{tab:N1333M}). 
%\todo{LWL: compare core sizes to beam?}
Note that we already excluded small, compact cores when selecting targets, and thus even the smallest cores identified in DiSCo have diameters $\gtrsim 14$\arcsec, or $1.5$ times of the $\sim 9$\arcsec{} beam.

Among the total of 66 targets in B1 and NGC\,1333, two appear to be not core-like in \NtwoH{} (see Fig.~\ref{fig:nocore}), both in NGC\,1333. There is basically no \NtwoH{} emission associated with protostellar core Per-emb-24 (Fig.~\ref{fig:nocore}, left), which could suggest that this protostar is actually at a later evolutionary stage instead of a Class 0 source \citep[see e.g.,][]{Tobin_VANDAM_2016}.
On the other hand, while being previously identified as a core using GAS data with 30\arcsec{} resolution, starless ``core'' N1333-2 is actually a ring-like structure (Fig.~\ref{fig:nocore}, right). These two targets were thus removed from the analysis.

In some cases, multiple protostars and/or local maxima of \NtwoH{} emission are too close to one another, and there is no closed contour (either $N_{\rm H}$ column density or \NtwoH{} emission) with a single target. This is the case for Per-emb-6/Per-emb-10, B1-bN/B1-bS (see Fig.~\ref{fig:B1core}), Per-emb-27/N1333-8, Per-bolo-45/N1333-21, Per-emb-12/Per-emb-13/IRAS4B' (see Fig.~\ref{fig:N1333core}), and the SVS\,13 region (see Fig.~\ref{fig:SVS13}). Also, when the second local maximum around the core area is weak or small in size, we loosen our criterion to include such substructures inside the identified core boundary. This is the case for N1333-1, N1333-13, N1333-17, N1333-22, and N1333-23 (see Figs.~\ref{fig:N1333core} and \ref{fig:SVS13}; the larger core area are marked with dark purple ellipses). All these cores are categorized as 'multiples' in our analysis.
%(see e.g.,~Figs.~\ref{fig:dI} and \ref{fig:JR}).

Interestingly, if we define the normalized intensity variation across the core to be 
\begin{equation}
    \frac{\Delta I}{I_{\rm peak}} \equiv \frac{I_{\rm peak} - I_{\rm boundary}}{I_{\rm peak}}\ \ \ {\rm or} \ \ \frac{\log N_{\rm peak} - \log N_{\rm boundary}}{\log N_{\rm peak}},
\end{equation}
where $I_{\rm boundary}$ is the intensity of the contour level corresponding to the core boundary, we find that the normalized intensity difference $\Delta I/I_{\rm peak}$ tends to have a power-law correlation with size, as shown in Fig.~\ref{fig:dI}. 
While both the log of $N_{\rm H}$ column density and \NtwoH{} emission were used as $I$ here,\footnote{\revt{Note that there is an offset in $\Delta I/I_{\rm peak}$ values derived from these two types of intensity, which is not surprising given that one is measured in linear scale and one in log scale.}} 
we see that they follow a similar power-law correlation to the core size, roughly as $\Delta I/I_{\rm peak} \propto r$. 
%This implies a core structure where the density distribution is relatively flat at small radii and drops off more steeply at larger radii. 
While neither \NtwoH{} emission intensity nor H$_2$ column density can be directly linked to gas density without further information of gas properties, we note that such correlation resembles a Bonner-Elbert sphere with a flat central region surrounded by a power-law drop-off in density distribution. 
This seems consistent with the expected power-law density profile of dense cores,
$\rho \propto r^{-p}$, 
%where $3/2\ {\rm (free\ fall)}\ \leq p \leq\ 2$ (thermal\ equilibrium),
from observations \citep[see e.g.,][]{Adams_Intensity_1991,WT_coreprof_1999,Shirley_cores_2000,Bacmann_cores_2000,Alves_cores_2001,Evans_cores_2001,Motte_Andre_2001, Kirk_SCUBAcores_2005}, theories (e.g.,~the singular-isothermal sphere in \citealt{Shu1987} and the Bonnor-Ebert sphere in \citealt{Ebert_1955} and \citealt{Bonnor_1956}), and simulations \citep[e.g.,][]{GO11,GO15,Offner_2022}.
%Note that generally the density profile of dense cores follows a power-law form as \citep[see e.g.,][]{Adams_Intensity_1991} $\rho \propto r^{-p}$, where $3/2\ {\rm (free\ fall)}\ \leq p \leq\ 2\ {\rm (thermal\ equilibrium)}$.
%For a locally flat core with $\rho \propto r^{-p}$ and depth $L$, the column density is $N \sim \rho \cdot 2\pi r dr \cdot L \propto r^{-p+2}$ .

%\begin{figure*}
%	\includegraphics[width=\textwidth]{Fig/ALLclass0_vlsr6.png}
%    \caption{All protostellar cores. }
%    \label{fig:protoV}
%\end{figure*}

%\begin{figure*}
%	\includegraphics[width=\textwidth]{Fig/ALLcores_vlsr_starless_5x.png}
%    \caption{All starless cores. }
%    \label{fig:coreV}
%\end{figure*}

%\subsection{Velocity Gradient and The $J-R$ correlation}
\subsection{\texorpdfstring{Velocity Gradient and The $J-R$ correlation}{Velocity Gradient and The J-R correlation}}
\label{sec:JR}

\begin{figure*}
	\includegraphics[width=\textwidth]{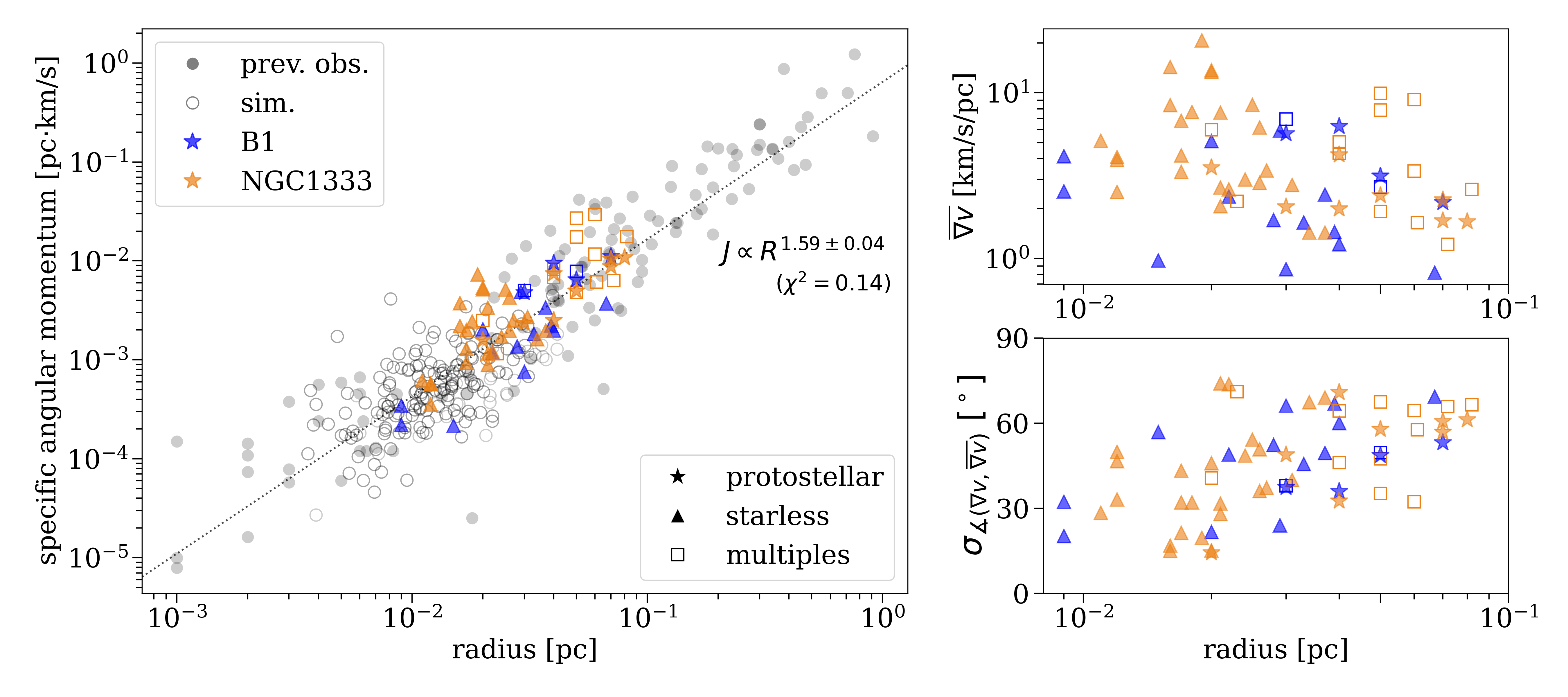}
    \caption{{\it Left:} The $J-R$ correlation combining DiSCo data and previous observations ({\it solid circles}) and simulations ({\it open circles}). DiSCo objects for B1 and NGC\,1333 regions are marked by different colors, and various symbols represent the three categories: protostellar cores ({\it stars}), starless cores ({\it triangles}), and multiples ({\it squares}). 
    %\revt{Note that the slope $J\propto R^{1.5}$ is simply a theoretical reference; the actual fit of the power-law index using DiSCo data is $1.55\pm 0.15$, and is $1.59\pm 0.04$ when including all data on the plot.}  
    \revtt{Dotted line shows the least-square fit using all data on the plot ($J\propto R^{1.59\pm 0.04}$ with $\chi^2 = 0.14$), which is very close to the fit using DiSCo data only ($J\propto R^{1.55\pm 0.15}$ with $\chi^2 = 0.09$).}
    {\it Right:} The mean velocity gradient ({\it top panel}) and the dispersion of the velocity gradient direction ({\it bottom panel}) within the core, as functions of the core size. }
    \label{fig:JR}
\end{figure*}

Following our pilot study, we calculate the gradient of $v_{\rm lsr}$ \revt{locally} at each map pixel first \revt{by averaging the difference of $v_{\rm lsr}$ among nearest neighbors, then take the vector average of $\nabla v_{\rm lsr}$ to derive the mean gradient direction (the arrows in Figures~\ref{fig:B1core}--\ref{fig:N1333core}). The length of this mean vector is then}  
%average over the defined core region to derive 
the mean linear velocity gradient at core scale, $\overline{\nabla v}$.
Assuming the observed velocity gradient is due to core rotation, the angular velocity of the core is simply $\omega = \overline{\nabla v}$ \citep{Goodman1993},
which can then be combined with the fitted core radius $R$ to derive the specific angular momentum $J \equiv L/M \sim R^2\omega = R^2 \overline{\nabla v}$. 
We then compare the $J-R$ relation from our DiSCo data with previous observations \citep{Goodman1993,Caselli2002,XChen_2007,Pirogov_2003,Tobin_2011,Yen15_B335disk} and simulations \citep{CO18}, which is shown in Fig.~\ref{fig:JR} (left panel). 

Clearly, DiSCo cores blend nicely in with the existing trend \revt{that roughly follows} a power-law correlation $J\propto R^{1.5}$ \citep[see e.g.,][]{GBTcore19}.
\revt{
The least-square fit of DiSCo data gives $J\propto R^\alpha$ with $\alpha = 1.55\pm 0.15$, and $\alpha = 1.59 \pm 0.04$ when combining all data (from both observations and simulations) shown in Figure~\ref{fig:JR} (left panel).
}
%\footnote{\revt{Note that we chose not to fit the data to measure the power-law index because we think the trend is visually obvious enough. Plus, while the trend is obvious, the scatter is still large (see text), which makes a fit less meaningful. We therefore think it is more insightful to provide a reference line from theoretical model to guide the eye instead of giving a numerically fitted value with high uncertainty. }} 
However, we note that while the core assembly roughly follows the same power law, the scatter in $J$ for a given radius is pretty large, more than one order of magnitude. Considering the definition of $J\equiv R^2 \overline{\nabla v}$ already guarantees sharp dependence on $R$, this suggests that the dependence of $\overline{\nabla v}$ on core size is relatively weak and has large scatter. We demonstrate this in the upper-right panel of Fig.~\ref{fig:JR}, which plots the radial dependence of $\overline{\nabla v}$. This shows that the mean velocity gradient within individual cores $\overline{\nabla v}$ has a noisy negative correlation with the core radius, which is the main source of uncertainty in the $J \propto R^\alpha$ power law dependence.
%See Sec.~\ref{sec:rotation} below for further discussions.

\section{Discussions}

\subsection{Solid-body Rotation?}
%\label{sec:discussion}
%\subsection{Solid-body rotation?}
\label{sec:rotation}

%\todo{- more elaboration on how widely the solid-body rotation is assumed and the implication of this being wrongly assumed}

The protostellar evolution of angular momentum at core scale is critical in shaping the follow-up formation of disks and/or multiple systems \citep{OffnerPPVII}, and a rotating dense core (plus turbulence as velocity perturbations) is a generally-adopted initial condition in numerical studies \citep[e.g.,][or see references in the recent reviews by \citealt{Pineda_PPVII} and \citealt{Zhao_review_2020}]{Seifried_disk_2013, Lam_Disk_2019, Tsukamoto_disk_2020}.
The most direct observable feature of solid-body rotation would be a uniform gradient of line-of-sight velocity across the core, as discussed in \cite{Goodman1993}. The derivation of the specific angular momentum $J \equiv L/M = R^2\nabla v$ is also based on such an assumption. 
However, as previous observations also reported, only very few dense cores show clear, monotonic velocity gradient within them \citep[see e.g.,][]{Caselli2002}.
\revt{
Recent observations by \cite{Pineda_jr_2019} also revealed features of non-solid-body rotation within dense cores.
}
Such deviation from the idealized assumption of solid-body rotation could be the result from turbulent motion at the core scale \citep[see e.g.,][]{CO18}.

To investigate such internal motion, we measured the dispersion of line-of-sight velocity gradient directions, $\sigma_{\measuredangle (\nabla v, \overline{\nabla v})}$, 
%as 
%\begin{equation}
%    \sigma_{\measuredangle (\nabla v, \overline{\nabla v})} = \sqrt{\overline{\left(\Delta\measuredangle(\nabla v, \overline{\nabla v})\right)^2}}
%\end{equation}
where $\measuredangle(\mathbf{x},\mathbf{y})$ represents the angle between the two vectors $\mathbf{x}$, $\mathbf{y}$.
%where $N$ is number of image pixels considered in the calculation.
This quantity is listed as the uncertainty of $\theta_{\overline{\nabla v}}$ in Tables~\ref{tab:B1}--\ref{tab:N1333M}, which indicates whether or not the mean velocity gradient within the core is well-defined. By plotting it against core size (Fig.~\ref{fig:JR}, bottom right), we see that larger cores tend to have more chaotic internal motion (i.e., larger dispersion in velocity gradient direction). This could be due to the turbulence cascade from the parent cloud. However, we also note that these larger, more turbulent cores are mostly protostellar cores and/or multiples, and thus their velocity fields could have been distorted by gravitational collapse or the interaction between the other cores.

\begin{figure*}
	\includegraphics[width=\textwidth]{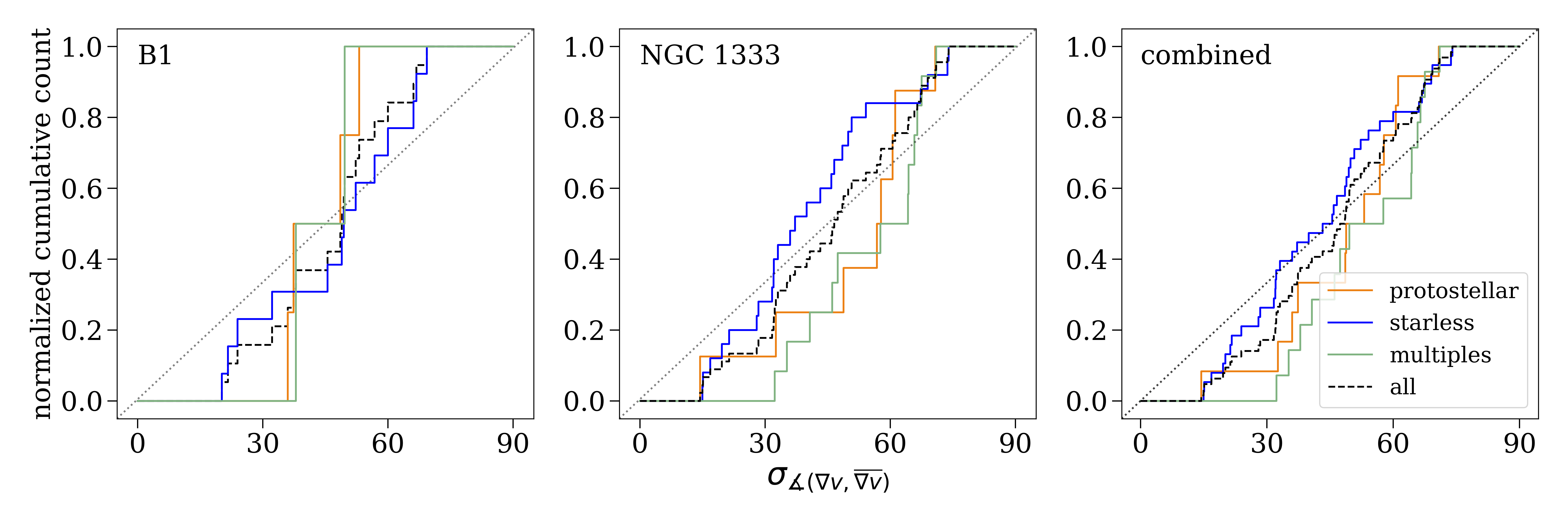}
    \caption{\revt{Cumulative distributions} of the dispersion of gradient velocity direction for different types of cores ({\it different colors; see legend}) and for separate regions ({\it left and middle}) and all regions combined ({\it right}). Most cores tend to have a dispersion of $\sim 30-60^\circ$ in the direction of the velocity gradient, indicating there is no prominent overall velocity gradient direction. }
    \label{fig:gradV}
\end{figure*}

We further investigate the internal velocity variation for cores from different environments and at different evolutionary stages. Fig.~\ref{fig:gradV} shows the step-style histograms of the dispersion of velocity gradient direction, $\sigma_{\measuredangle (\nabla v, \overline{\nabla v})}$, within starless (blue) and protostellar (orange) cores, as well as those with multiple cores (green), for both B1 (left panel) and NGC\,1333 (middle panel) regions, and for the two regions combined (right panel). For rigid-body rotation, there should be a prominent direction of velocity gradient within the core, and thus the angle dispersion of $\nabla v$ should be small. This is clearly not the case for the DiSCo targets presented here, as most of the cores appear to have angle dispersion $\sim 30-60^\circ$ in their velocity gradient direction. Interestingly, there are clearly more starless cores with small ($\lesssim 30^\circ$) dispersion in $\measuredangle(\nabla v, \overline{\nabla v})$ than protostellar cores, which makes sense because more severe, global collapse may have started in protostellar cores, and thus the velocity structure would not be as smooth as in early-stage starless cores. 
%\todo{- some discussions on multiples}
We also note that, while there are not many samples, cores with multiple emission peaks (``multiples'' in the plot) do not seem to differ significantly from protostellar cores, which is generally consistent with recent theoretical work by \cite{Smullen_multi_2020} that there seems to be very little correlation between core properties and multiplicity.

The large dispersion in velocity gradient direction indicates that solid-body rotation may not be a good assumption for core-scale gas motion.
More importantly, we would like to emphasize that even if the cores show prominent, uniform velocity gradient direction, it is not guaranteed that such velocity gradient is due to rotation
(\citealt{CO18}, also see e.g.,~\citealt{ChenFilament2020,HsiehFilament21} for related discussions in filaments). 
To test this, we follow the analysis conducted by \cite{CO18} and measure the angle between the fitted core major axis $a_{\rm core}$ and the averaged velocity gradient direction. If the velocity gradient across the core is due to rotation, we expect it to be roughly parallel to the core's major axis, because the classical theory of star formation envisioned the core to flatten along the rotational axis \citep[see e.g.,][]{Shu1987}.
The results are illustrated in Fig.~\ref{fig:rot}, which shows the \revt{cumulative distributions} of the rotation-core alignment angle, $\measuredangle (a_{\rm core}, \overline{\nabla v})$. Note that we limit this angle to be within $[0^\circ,90^\circ]$. We see that there is no clear correlation between the core orientation (as measured by the major axis) and the velocity gradient direction, as the step functions are all closer to the diagonal line (complete random distribution). 
This is a strong evidence suggesting that the velocity gradients measured within dense cores are probably not \revt{pure rotation but a mix between turbulence and real rotation}, in good agreement with the simulation results in \cite{CO18}.

\begin{figure}
    \includegraphics[width=\columnwidth]{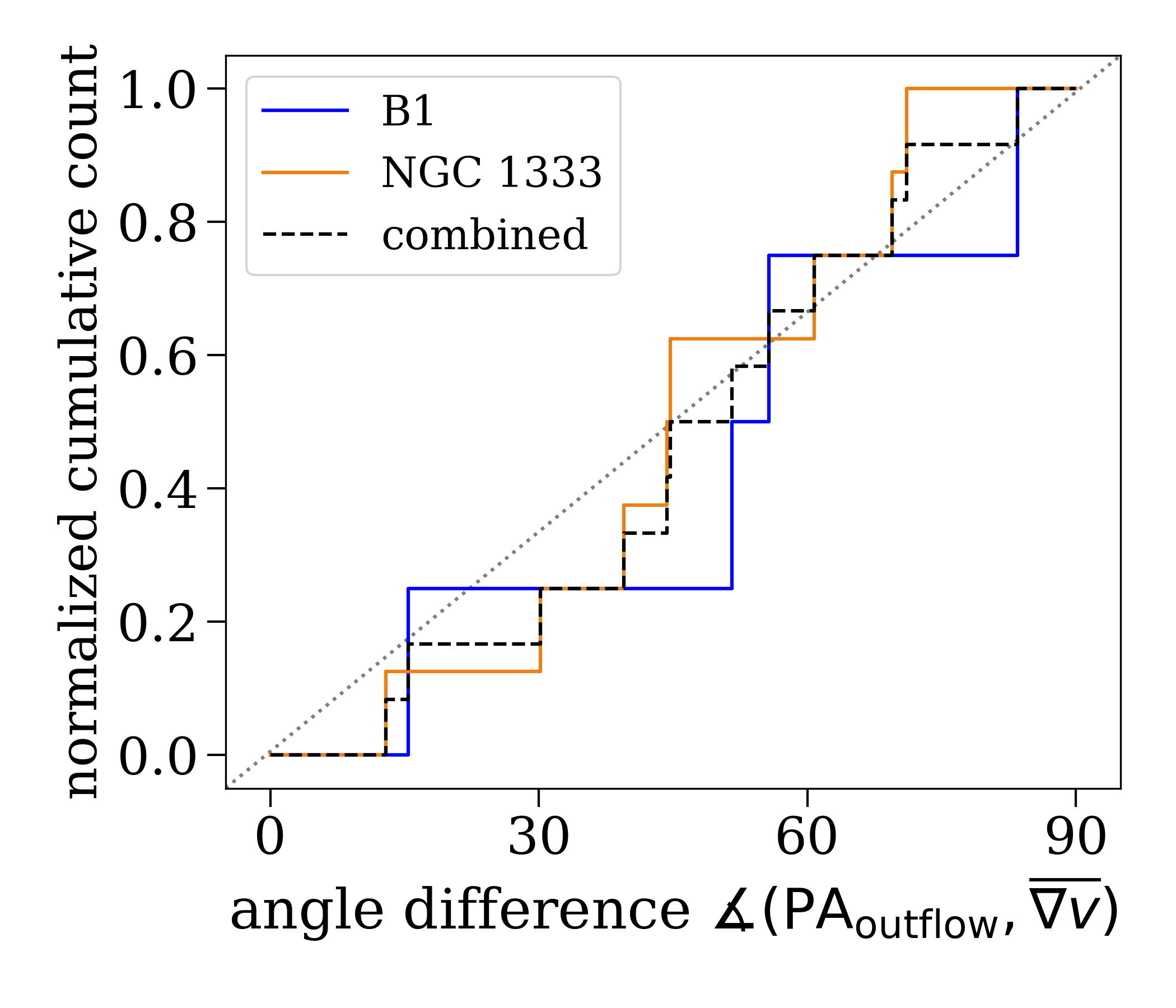}
    \caption{\revt{Cumulative distribution} of the alignment between the protostellar outflows measured in MASSES \citep{Stephens_MASSES_2018}) and the derived mean velocity gradient $\overline{\nabla v}$ within corresponding DiSCo cores. Our results suggest there is little or no correlation between the velocity structure at core scale and the outflow orientation, with a slight preference toward a perpendicular alignment between these two, which is the case described in the classical theory \citep[e.g.,][]{Shu1987}.}
    \label{fig:OFPA}
\end{figure}

\begin{figure*}
	\includegraphics[width=\textwidth]{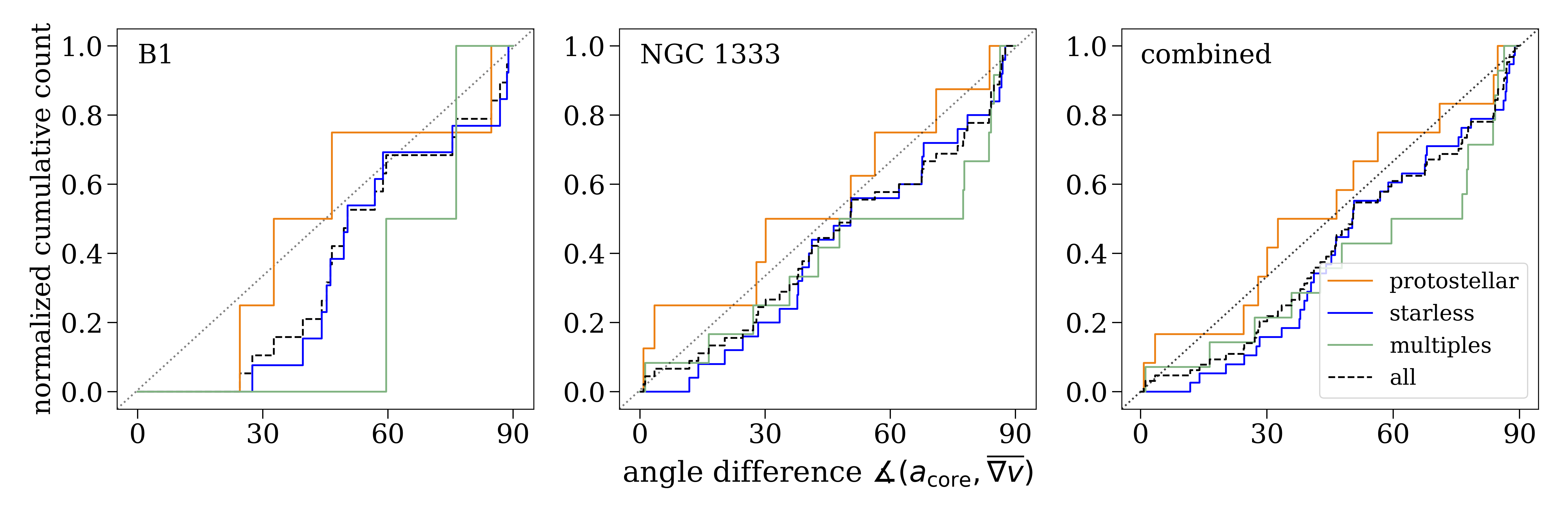}
    \caption{\revt{Cumulative distributions} of the rotation alignment, $\measuredangle (a_{\rm core}, \overline{\nabla v})$, for protostellar ({\it orange}), starless ({\it blue}), and multiple ({\it green}) cores as well as all cores combined ({\it black dashed lines}) for both B1 ({\it left panel}) and NGC\,1333 ({\it middle panel}) regions and the two regions combined ({\it right panel}). Most of the step functions are very close to that of a random distribution ({\it gray dotted diagonal line}), which suggests that there is no preferred alignment between the core and the velocity gradient direction within it. }
    \label{fig:rot}
\end{figure*}

%\todo{- new figure on outflows}
In addition, we compared the velocity gradient directions in our protostellar core targets to their outflow directions reported in MASSES \citep{Stephens_MASSES_2018}. The results are shown as \revt{cumulative distributions} in Fig.~\ref{fig:OFPA}. 
There seems to be no correlation between the mean velocity gradient direction and the outflow orientation, and thus the angle between these two directions has a relatively random distribution. 
This is again in contrast to the idealized assumption of solid-body rotation, because the protostellar outflows tend to be launched along the rotational axis of the core, and thus the observed outflows are expected to be aligned perpendicular to the core-scale velocity gradient direction. 

If solid-body rotation is not a reasonable assumption, the so-called specific angular momentum $J\equiv L/M$ is in fact simply $R^2 \nabla v = R\Delta v$, where $\Delta v = \nabla v \cdot R$ is the velocity difference across the core. The $J-R$ correlation discussed in Sec.~\ref{sec:JR}, $J \propto R^{1.5}$, thus simply reduces to $\Delta v (R) \propto R^{0.5}$. As noted in \cite{CO18}, this resembles the Larson's law of cloud-scale turbulence, $\sigma_v (\ell) \propto \ell^{0.5}$. 
Our results therefore support the hypothesis \citep[see e.g.,][]{CO18,BB_turb_2000} that the observed velocity gradient across dense core has a tight connection to the larger-scale gas dynamics instead of local rotation.
\revt{
For example, Figures~\ref{fig:B1result} and \ref{fig:N1333result} show that the gas dynamics are spatially coherent (i.e., continuous and smooth between cores). This suggests that when we measure the velocity gradient within cores, we may be simply taking samples of the gas dynamics at the larger scale. If the large-scale gas dynamics are directly related to cloud-scale turbulence, this could be a hint of the turbulence origin of core-scale velocity gradient. More detailed investigations are needed to provide better, quantitative evidences.
}

\subsection{Connecting to Larger Scales}

\revt{
While the main focus of DiSCo is on dense core dynamics at $0.01$\,pc scale, the full-range coverage of physical scale and the large field of view ($\sim 2$\arcmin $\times 2$\arcmin ) of GBT-Argus also allows DiSCo observations to recover gas dynamics at cloud ($0.1-1$\,pc) scales. The comparison between DiSCo and previous studies with similar coverages but lower resolutions therefore could provide great insight into the evolution of gas dynamics across different scales during star formation.

The two surveys on which our target selection was based, CLASSy and GAS, provided the most direct comparison to our DiSCo data. While a detailed study on connecting DiSCo results to GAS results is still ongoing, we have compared \NtwoH{} data from our pilot study \citep{GBTcore19} to the corresponding CLASSy regions, and concluded that the velocity structure probed by GBT-{\it Argus} is roughly consistent with CLASSy results. We note that one caveat of such comparison is that the \NtwoH{} data from CLASSy have $\sim 5\times$ worse spectral resolution than GBT-{\it Argus}, which may affect the spectral fitting result and the subsequent velocity analysis. 

In addition to the aforementioned CLASSy and GAS, \cite{Dhabal_N1333_2019} shows high-resolution NH$_3$ maps of NGC\,1333 using GBT and VLA combined. By looking at the DiSCo \NtwoH{} velocity map of NGC\,1333 in Figure~\ref{fig:N1333result} (right panel) and the GBT+VLA NH$_3$ velocity map (Figure~8 in \citealt{Dhabal_N1333_2019}), one can tell that these two datasets are highly consistent. While detailed, quantitative comparison is beyond the scope of this work, we stress that our DiSCo data also shows a huge velocity difference (Figure~\ref{fig:N1333result} (right panel)) across the SVS\,13 region toward IRAS\,4 region in NGC\,1333 (see Figure~8 in \citealt{Dhabal_N1333_2019} for definitions), which is consistent with the turbulent cell colliding region as described in \cite{Dhabal_N1333_2019}. 

Last but not the least, \cite{Hacar_N1333_2017} conducted \NtwoH{} observations using the IRAM 30m telescope (30\arcsec{} resolution) and identified velocity-coherent structures (termed ``fibers’’) in the main region of NGC\,1333. By looking at dense core locations from Figure~\ref{fig:N1333result} (left panel) and comparing with Figure~13 in \cite{Hacar_N1333_2017}, there seems to be a trend for cores to form around the intersection of fibers. This is consistent with the results reported in \cite{HopeGAS19}, who identified pressure-bound cores using GAS data and pointed out that dense cores tend to form at the turning point along the velocity axis in position-position-velocity space (i.e., local extremes). Further studies are needed to provide more insight into this core-fiber connection.
}

\section{Summary}
\label{sec:sum}

We present the first results from the DiSCo project, which is a GBT-{\it Argus} Large Program to survey the velocity structures within all starless and Class~0 cores in Perseus using the \NtwoH{} J=1-0 line. 
Data from the B1 and NGC\,1333 regions in Perseus are reported here.
We characterize core boundaries using either the {\it Herschel} column density map (most of the protostellar cores) or the integrated \NtwoH{} emission (starless cores) from this project. Line-of-sight velocity is measured by fitting all hyperfine structures of \NtwoH{} J=1-0 altogether. 

We further calculate the gradient vector of the line-of-sight velocity within cores. The velocity gradient amplitude is then used to derive the specific angular momentum assuming it originates from core-scale rotation.
Our results are highly consistent with previous observations and simulations; 
the specific angular momentum $J$ roughly follows a power-law correlation with the core radius $R$, $J\propto R^\alpha$ where
\revt{
$\alpha \approx 1.55\pm 0.15$ according to a least-squares fit.
}

On the other hand, we use the velocity gradient direction to examine whether or not the core-scale velocity structure can be best explained by solid-body rotation. Since most of the DiSCo cores show large dispersion in the velocity gradient direction ($\gtrsim 30^\circ$), this suggests that the mean velocity gradient (that we used to calculate the specific angular momentum) is not truly representative,
Hence, the community needs to be careful when adopting the linear approximation on core-scale velocity structure, especially on data with insufficient spatial or spectral resolution.
Also, the velocity gradient direction shows no obvious correlation with either the core orientation or protostellar outflow direction,
which means the velocity structure within cores may have external origins.

The key scientific goal of the DiSCo project is to produce finely-resolved velocity information at dense core scale and use it to provide justifications on theoretical models of the early stages of protostellar evolution. 
\revt{
When complete, DiSCo will have high-quality velocity information on more than 100 cores. This sample will provide strong statistical evidence about whether rotation is a common feature of dense cores, and whether the core-scale dynamics are generally consistent with cloud-scale motions. 
}
While this is not the complete data set and more detailed investigations are needed to better utilize the spectral resolution and dynamic range of the DiSCo data, the simple analysis conducted here shows evidence supporting the turbulent origin of the core-scale velocity structure.

\section*{Acknowledgements}

%The Acknowledgements section is not numbered. Here you can thank helpful colleagues, acknowledge funding agencies, telescopes and facilities used etc. Try to keep it short.
\revtt{We thank the referee for detailed and thorough reports.}
The authors would like to thank the {\it Argus} instrument team from the Stanford University, Caltech, JPL, University of Maryland, University of Miami, and the Green Bank Observatory for their efforts on the instrument and software that have made this work possible. The {\it Argus} instrument construction was funded by the National Science Foundation (NSF) ATI-1207825. Green Bank Observatory is a facility of the National Science Foundation and is operated by Associated Universities, Inc. 
CYC and ZYL acknowledge support from NSF grant AST-1815784. 
ZYL is supported in part by NASA 80NSSC20K0533 and NSF AST-2307199.
SSRO was supported by the NSF through CAREER award 1748571, AST-2107340 and AST-2107942 and by the Oden Institute through a Moncrief Grand Challenge award.
This research made use of \texttt{Astropy} (\url{http://www.astropy.org}), a community-developed core \texttt{Python} package for Astronomy \citep{astropy:2013, astropy:2018}. 
This work was performed under the auspices of the U.S. Department of Energy (DOE) by Lawrence Livermore National Laboratory under Contract DE-AC52-07NA27344 (CYC). LLNL-JRNL-843246-DRAFT. 

%%%%%%%%%%%%%%%%%%%%%%%%%%%%%%%%%%%%%%%%%%%%%%%%%%
\section*{Data Availability}

%The inclusion of a Data Availability Statement is a requirement for articles published in MNRAS. Data Availability Statements provide a standardised format for readers to understand the availability of data underlying the research results described in the article. The statement may refer to original data generated in the course of the study or to third-party data analysed in the article. The statement should describe and provide means of access, where possible, by linking to the data or providing the required accession numbers for the relevant databases or DOIs.
The DiSCo data will be available in the Green Bank Observatory Data Archive once the system is ready. 
Before that, 
all data underlying this article will be shared on reasonable request to the corresponding author.
%The data underlying this article are available in [repository name, e.g. the Dryad Digital Repository], at https://dx.doi.org/[doi]



\bibliographystyle{mnras}
\bibliography{reference} % if your bibtex file is called example.bib

%\begin{thebibliography}{99}
%\end{thebibliography}

%%%%%%%%%%%%%%%%%%%%%%%%%%%%%%%%%%%%%%%%%%%%%%%%%%

%%%%%%%%%%%%%%%%% APPENDICES %%%%%%%%%%%%%%%%%%%%%

%\appendix


%If you want to present additional material which would interrupt the flow of the main paper, it can be placed in an Appendix which appears after the list of references.

%%%%%%%%%%%%%%%%%%%%%%%%%%%%%%%%%%%%%%%%%%%%%%%%%%

% Don't change these lines
\bsp	% typesetting comment
\label{lastpage}
\end{document}